\documentclass[11pt]{article}
\usepackage{axodraw}
\usepackage{epsfig}
\usepackage{amsfonts}
\usepackage{amsmath}
\usepackage{bm,bbm}
\usepackage{cite}
\usepackage{jheppub}
\allowdisplaybreaks[1]

\input pix.sty

\newcommand{\mZ}{m_\rmii{$Z$}}
\newcommand{\rhoV}{\rho_\rmii{$V$}}
\newcommand{\rhoH}{\rho_\rmii{$H$}}
\newcommand{\rhoT}{\rho_\rmii{$T$}}
\newcommand{\rhoL}{\rho_\rmii{$L$}}
\newcommand{\lnf}{l^{ }_\rmi{1f}}
\newcommand{\lif}{l^{ }_\rmi{2f}}
\newcommand{\ltf}{l^{ }_\rmi{3f}}
\newcommand{\ko}{\omega} % {k_0}
\newcommand{\km}{k_-}
\newcommand{\kp}{k_+}

\renewcommand{\eq}{eq.~}
\renewcommand{\eqs}{eqs.~}
\renewcommand{\se}{sec.~}
\renewcommand{\ses}{secs.~}
\renewcommand{\fig}{fig.~}
\renewcommand{\figs}{figs.~}

\newcommand{\tinymsbar}{{\overline{\mbox{\tiny\rm{MS}}}}}
\newcommand{\Lambdamsbar}{{\Lambda_\tinymsbar}}

\newcommand{\alphas}{\alpha_{\rm s}}
\newcommand{\Nf}{N_{\rm f}}
\newcommand{\Nc}{N_{\rm c}}

\newcommand{\Tc}{T_{\rm c}}

\newcommand{\mE}{m_\rmii{E}}
\newcommand{\gE}{g_\rmii{E}}

\newcommand{\rmO}{{\mathcal{O}}}
\newcommand{\bmu}{\bar\mu}
\newcommand{\CA}{\Nc}
\newcommand{\CF}{C_\rmii{F}}

\def\lsi{\raise0.3ex\hbox{$<$\kern-0.75em\raise-1.1ex\hbox{$\sim$}}}
\def\gsi{\raise0.3ex\hbox{$>$\kern-0.75em\raise-1.1ex\hbox{$\sim$}}}
\newcommand{\lsim}{\mathop{\lsi}}
\newcommand{\gsim}{\mathop{\gsi}}

\newcommand{\sign}{\mathop{\mbox{sign}}}
\newcommand{\nF}{n_\rmii{F}}
\newcommand{\nB}{n_\rmii{B}}

\newcommand{\rmii}[1]{{\mbox{\tiny\rm{#1}}}}

\newcommand{\im}{\mathop{\mbox{Im}}}

\newcommand{\Tint}[1]{{\hbox{$\sum$}\!\!\!\!\!\!\!\int\,}_{\!\!\!\!\raise-0.9ex\hbox{$\scriptstyle{#1}$}}}
\newcommand{\Tinti}[1]{{{\Sigma}\!\!\!\!\raise0.3ex\hbox{$\int$}_\rmii{${#1}$}}}

\newcommand{\bi}{\begin{itemize}}
\newcommand{\ei}{\end{itemize}}
\newcommand{\hide}[1]{ }

\makeatletter \@addtoreset{equation}{section} \makeatother
\renewcommand{\theequation}{\arabic{section}.\arabic{equation}}
\makeatletter
\renewcommand\section{\@startsection {section}{1}{\z@}%
                                   {-5.5ex \@plus -1ex \@minus -.2ex}% bfr-
                                   {2.3ex \@plus.2ex}%
                                   {\normalfont\large\bfseries}}
\renewcommand\subsection{\@startsection{subsection}{2}{\z@}%
                                     {-3.25ex\@plus -1ex \@minus -.2ex}%
                                     {1.5ex \@plus .2ex}%
                                     {\normalfont\normalsize\bfseries}}
\renewcommand\thesection {\@arabic\c@section}
\renewcommand\thesubsection   {\thesection.\@arabic\c@subsection}
\renewcommand{\@seccntformat}[1]{%
\csname the#1\endcsname.\hspace{1.0em}}
\makeatother
\begin{document}

\flushbottom

%\renewcommand{\thefootnote}{\fnsymbol{footnote}}

%x \begin{titlepage}

\begin{flushright}
% DRAFT \\ 
December 2019 
% arXiv:1910.09567
\end{flushright}

\vspace*{0.5cm}

\title{Testing thermal photon and dilepton rates}

\author{G.~Jackson,}
\author{M.~Laine}

\affiliation{%
AEC, 
Institute for Theoretical Physics, 
University of Bern, \\ 
Sidlerstrasse 5, CH-3012 Bern, Switzerland}

\emailAdd{jackson@itp.unibe.ch}
\emailAdd{laine@itp.unibe.ch}

\abstract{%
We confront the thermal NLO vector spectral function (both the transverse 
and longitudinal channel with respect to spatial momentum, both above and 
below the light cone) with continuum-extrapolated lattice data 
(both quenched and with $N^{ }_{\rm f} = 2$, at $T \sim 1.2 \Tc^{ }$). 
The perturbative side incorporates new results, whose main features are 
summarized. The resolution of the lattice data is good enough to constrain
the scale choice of $\alpha^{ }_{\rm s}$ on the perturbative side. 
The comparison supports the previous indication that the true spectral 
function falls below the resummed NLO one in a substantial frequency domain. 
Our results may help to scrutinize direct spectral reconstruction attempts 
from lattice QCD. 
}

%% %\noindent
%% %PACS numbers: 
%% %11.10.Wx, %        Finite temperature field theory
%% { %11.15.Ha, %        Lattice gauge theory } 
%% %12.38.Bx, %        Perturbative calculations in QCD
%% %12.38.Mh, %        Quark--gluon plasma
%% %14.40.Nd, %        Bottom mesons
%% %\\

\keywords{Thermal Field Theory, Quark-Gluon Plasma, 
Lattice QCD, Perturbative QCD}
 
\maketitle

%x \end{titlepage}

% \renewcommand{\thefootnote}{\arabic{footnote}}
% \setcounter{footnote}{0}

%%%%%%%%%%%%%%%%%%%%%%%%%%%% SECTION %%%%%%%%%%%%%%%%%%%%%%%%%%%%%%%%%%%%
%
\section{Introduction}
\la{se:intro}

Photons and lepton-antilepton pairs produced in a heavy ion collision 
are experimentally measurable~(cf.,\ e.g.,\ refs.~\cite{exp1,exp2,exp3})
and, given that they do not interact after production,  
offer for a probe of the inner dynamics of strong interactions
in this environment. To leading order in the electromagnetic
fine-structure constant $\alpha^{ }_\rmi{em}$, 
the thermal parts of both production rates can be related to 
the spectral function $\rhoV^{ }$, associated with the QCD
vector current~\cite{old1,old2,old3},  
\ba
 \frac{{\rm d}\Gamma_\gamma({k})}{{\rm d}^3\vec{k}}
  & = &
 \frac{\alpha^{ }_\rmi{em}\,\nB^{ }(k) }{2 \pi^2 k}
 \sum_{i = 1}^{\Nf} Q_i^2 \;  
 {
  {\rhoV^{ }(k,{k})} 
 }
 \; + \; \rmO(\alpha_\rmi{em}^2) 
 \;, \la{photon}
 \\
%%%
 \frac{{\rm d} \Gamma_{\ell\bar\ell}\,(\omega,k)}
   {{\rm d}\omega\, {\rm d}^3 \vec{k}} 
  & \approx &
 \frac{\alpha_\rmi{em}^2  \nB^{ } (\omega)  } 
  {3 \pi^3 M^2} 
 \sum_{i = 1}^{\Nf} Q_i^2 \;  
 \rhoV^{ }(\omega,{k})
 \; + \; \rmO(\alpha_\rmi{em}^3) 
 \;.\la{dilepton}
\ea 
Here $\nB^{ }$ is the Bose distribution; 
$M \equiv \sqrt{\omega^2 - k^2}$, 
$\omega$ and $k$ are the invariant mass, energy, and 
momentum, respectively, of a virtual photon; 
$Q^{ }_i$ is the charge
of a quark of flavour $i$ in units of the elementary charge; 
disconnected contributions proportional to $(\sum_i Q^{ }_i)^2$
have been omitted;  
and we have simplified \eq\nr{dilepton} by considering 
energies $2 m^{ }_\ell \ll M \ll \mZ^{ }$.

There is a long history of perturbative 
determinations of $\rhoV^{ }$ in various kinematic domains. 
Focussing first on massless quarks, 
a next-to-leading order (NLO) computation 
at vanishing momentum ($k=0$) initially suggested
that perturbation theory works well~\cite{bps,ggp,aa}. 
However, pushing the energy towards a soft
regime ($\omega \ll \pi T$, $k=0$) 
and implementing Hard Thermal Loop (HTL) 
resummation, a large enhancement was found~\cite{htl,gelis1}. 
Subsequently the focus shifted to the more typical 
hard momenta ($k \sim \pi T$), where a logarithmic singularity, 
shielded by HTL-resummation, 
was identified when approaching the light cone
($M \ll \pi T$)~\cite{kls,rb,ar}. 
In addition, there are non-logarithmic terms 
of similar magnitude~\cite{gelis2}, 
originating from amongst others multiple scatterings with collinear
enhancement (the so-called LPM effect~\cite{gelis3}), whose systematic handling
necessitated a major effort~\cite{amy1,amy2,agmz,mg}. 
By now these resummed results have been extended up to
NLO close to the light cone~\cite{nlo1,nlo2}.
With different methods, 
the NLO level has also been reached above the light cone
($M\sim \pi T$)~\cite{master,nlo}, 
and the corresponding results have been shown to permit for a smooth
interpolation towards the light-cone ones~\cite{interpolation}.  
Far above the light cone, 
the spectral function is considerably simpler~\cite{simon}, 
and can in fact be determined to a high precision~\cite{cond}, by making
use of N$^4$LO vacuum results~\cite{baikov,baikov2}.
Finally, quark mass effects have been included up to the NLO level
at finite temperature, 
both for $m \gg \pi T$~\cite{mass} and for $m\lsim \pi T$~\cite{yannis}.  

Diverse as the progress is, it should be clear that eventually we need 
to go beyond perturbation theory in the determination of $\rhoV^{ }$. 
Lattice QCD entails the measurement of an imaginary-time correlation
function $ G^{ }_\rmii{$V$}(\tau,{k}) $, which is related to 
$\rhoV^{ }$ through 
\be
 G^{ }_\rmii{$V$}(\tau,{k})
 = 
 \int_0^\infty \! \frac{{\rm d}\omega}{\pi}
 \, \rhoV^{ }(\omega,{k}) \, 
 \frac{\cosh [\omega (\frac{\beta }{2}-\tau) ] }
 {\sinh [ \frac{\omega \beta }{2} ] }
 \;, \quad
 \beta \equiv \frac{1}{T}
 \;. \la{relation}
\ee
The inversion of this relation is notoriously challenging
(cf.,\ e.g., ref.~\cite{rev}). 
A recent attempt was made in ref.~\cite{photon}, 
for continuum-extrapolated quenched QCD. It is clear from 
\eq\nr{relation} that, apart from the physical domain $\omega > k$, 
lattice results are also affected by the spacelike domain $\omega < k$.
However it can be argued 
that, in infinite volume, $\rhoV^{ }$ should be smooth across
the light cone~\cite{qhat}. Thus ref.~\cite{photon} made use of 
perturbative information at $M \gsim \pi T$ and a fitted 
interpolating polynomial 
at $0 \le \omega \lsim \sqrt{k^2 + (\pi T)^2}$. A subsequent work 
considered $\Nf = 2$ data~\cite{harvey}, noting that
for the photon channel the contribution
of a longitudinal polarization can be subtracted and replacing
the interpolating polynomial through a Pad\'e ansatz.
Further ideas at implementing analytic continuation
have also been put forward~\cite{screening,hbm}.  

The purpose of the present paper is to scrutinize
the spectral reconstructions of refs.~\cite{photon,harvey}. 
With this aim we improve the status of perturbative predictions 
in two respects: we incorporate 
full NLO results for $\omega < k$~\cite{gj}, 
and consider separately the transverse
and longitudinal polarizations 
as proposed in ref.~\cite{harvey}. 
After implementing 
proper resummation close to light cone, these expressions
can be inserted on the right-hand side of 
\eq\nr{relation}, and subsequently the left-hand side can be 
compared with lattice data. The perturbative results depend
on a parameter, namely the value of  
the renormalized gauge coupling, and these comparisons permit to 
``calibrate'' the choice made. 

Our presentation is organized as follows. 
In \se\ref{se:setup} we define the basic quantities considered. 
In \se\ref{se:limits} we consider various limits, theoretical
constraints, and resummations that pertain to their perturbative
determination. Comparisons with quenched and unquenched 
lattice data comprise \se\ref{se:lattice},
whereas conclusions are offered in \se\ref{se:concl}. 

%%%%%%%%%%%%%%%%%%%%%%%%%%%% SECTION %%%%%%%%%%%%%%%%%%%%%%%%%%%%%%%%%%%%
%
\section{Basic setup}
\la{se:setup}

Consider the Euclidean vector correlator
\be
 \Pi^{ }_{\mu\nu}(K)
 \; \equiv \; 
 -\, \int_X e^{i K\cdot X} 
 \Bigl\langle
   (\bar\psi \gamma^{ }_\mu \psi)(X) \, 
   (\bar\psi \gamma^{ }_\nu \psi)(0) 
 \Bigr\rangle^{ }_T
 \;, \la{Pi_munu}
\ee
where $K = (k^{ }_n,\vec{k})$, $X = (\tau,\vec{x})$, 
$\{\gamma^{ }_\mu,\gamma^{ }_\nu\} = 2 \delta^{ }_{\mu\nu}$, 
and $\langle ... \rangle^{ }_T$ denotes a thermal average on a volume
with a temporal extent $\tau \in (0,\beta)$. Correspondingly 
$k^{ }_n$ is a bosonic Matsubara frequency, 
{\it viz.}\ $k^{ }_n = 2\pi n T$, with 
$n \in \mathbbm{Z}$.
We denote $K^2 \equiv k_n^2 + k^2$, with $k \equiv |\vec{k}|$.
An overall minus sign has been inserted in \eq\nr{Pi_munu}
for later convenience. 

We are mostly interested in a spectral function, which can be obtained 
as an imaginary part of the Euclidean correlator, 
\be
 \rho^{ }_{\mu\nu}(\mathcal{K}) = 
 \im \bigl[ \Pi^{ }_{\mu\nu}(K)
     \bigr]^{ }_{k^{ }_n \to -i [\ko^{ } + i 0^+]}
 \;. \la{cut}
\ee
Its argument is the Minkowskian 
four-momentum $\mathcal{K} \equiv (\ko,\vec{k})$, with 
$\mathcal{K}^2 \equiv M^2$.

Following ref.~\cite{harvey}, we are particularly interested 
in the linear combinations
\be
 \rhoV^{ } \; \equiv \; 
 \rho^{ }_{\mu\mu}
 \;, \quad
 \rhoH^{ } \; \equiv \; \rhoV^{ } +
 \frac{(D-1) M^2}{k^2} \rho^{ }_{00}
 \;, \la{harvey}
\ee
where repeated indices are summed over. Here $D \equiv 4 - 2\epsilon$
is the dimension of spacetime. 
%% In vacuum, $\rhoH^{ }$ vanishes because of Lorentz symmetry. 
On the light cone, %% {\it viz.}\ $M^2 = 0$,
$\rhoV^{ }$ and $\rhoH^{ }$ coincide, so that we may replace
$\rhoV^{ }$ through $\rhoH^{ }$ in \eq\nr{photon}.   
At leading order (cf.,\ e.g.,\ ref.~\cite{ga}),  
\ba
 \rho^{ }_\rmii{V} & = &
  \frac{ \Nc M^2 }{4 \pi k }
  \Bigl\{
   2 T \bigl[ \lnf(\kp^{ }) - \lnf(|\km^{ }|)\bigr]
 + k \, \theta(\km^{ })
  \Bigr\}
 \;, \\
  \rho^{ }_{00} & = &
  - \frac{ \Nc }{12 \pi k }
  \Bigl\{ 
    24 T^3 \bigl[ \ltf(\kp^{ }) - \ltf(|\km^{ }|)\bigr]
% \nn 
% & & \hspace*{1.1cm} + \, 
 + 12 k T^2 \bigl[ \lif(\kp^{ }) + \sign(\km^{ })\, \lif(|\km^{ }|)\bigr]
 +  k^3 \, \theta(\km^{ }) 
  \Bigr\} 
 \;, \nonumber
\ea 
where we have defined 
$k^{ }_{\pm} \equiv (\ko^{ }\pm k)/2$ and introduced
the polylogarithms 
\be
 \lnf(\ko) \; \equiv \; \ln \Bigl( 1 + e^{-\ko/T} \Bigr)
 \;, \quad\;
 \lif(\ko) \;\, \equiv \; \mbox{Li}^{ }_2 \Bigl(-e^{-\ko/T}\Bigr)
 \;, \quad\; 
 \ltf(\ko) \;\, \equiv \; \mbox{Li}^{ }_3 \Bigl(-e^{-\ko/T}\Bigr)
 \;. \la{polylogs}
\ee

Denoting by $g^2 = 4 \pi \alphas^{ }$ the gauge coupling, 
by $\Nc^{ }$ the number of colours, 
by $\CF \equiv (\Nc^2-1)/(2\Nc)$ the quadratic Casimir coefficient,  
and by $\Tinti{\{ P \}}$  
a sum-integral with fermionic Matsubara momenta,
the NLO expressions for $\Pi^{ }_\rmii{$V$} \equiv \Pi^{ }_{\mu\mu}$
and $\Pi^{ }_{00}$ can be cast in the forms 
% ({\em use $\Delta^{ }_X \equiv X^2$??})
\ba
 \Pi^{ }_\rmii{$V$}(K) \!\! & = & \!\! 
 2 (D-2) \CA 
 \Tint{\{P\}} 
 \biggl[ \frac{2}{P^2} -  \frac{K^2}{P^2(P-K)^2}  \biggr]
%%%%%%%%%%%%%%%%%%%%%
 \nn 
 && \hspace*{-1.5cm} +\,
 4 (D-2) g^2 \CA \CF  
 \Tint{ \{PQ\} }
  \biggl\{
      \biggl[
      \frac{D-2}{P^4}
       - \frac{2}{P^2(P-K)^2}
       - \frac{(D-2)K^2}{P^4(P-K)^2}
      \biggr]
      \biggl[ 
        \frac{1}{Q^2} - \frac{1}{(Q-P)^2} 
      \biggr]
%%%%%%%%%%%%%%%%%%%%%
  \nn & & \;
  - \, \frac{D-4}{Q^2(Q-P)^2}
    \biggl[
     \frac{1}{P^2} - \frac{1}{ (P-K)^2}  
    \biggr]
  - \frac{\fr12\! (D-7) K^2}
  {P^2(P-K)^2Q^2(Q-K)^2}
%%%%%%%%%%%%%%%%%%%%%
  \nn[2mm] & & \; + \, 
  \frac{(D-6) K^2 - 2(D-2)K\cdot Q}
  {P^2 (P-K)^2 Q^2 (Q-P)^2}  
  + \frac{K^4}
  {P^2 (P-K)^2 Q^2 (Q-K)^2 (Q-P)^2 }
  \biggr\} 
 \;, \la{rhoV} \\[2mm] 
%%%%%%%%%%%%%%%%%%%%%%%%%%%%%%%%%%%%%%%%%%%%%%%%%
 \Pi^{ }_{00}(K) \!\! & = & \!\! 
 2 \CA 
 \Tint{\{P\}} 
 \biggl[ 
 \frac{2}{P^2} 
 - 
\frac{K^2 + 4 p^{ }_n (p^{ }_n - k^{ }_n)}{P^2(P-K)^2}
 \biggr]
%%%%%%%%%%%%%%%%%%%%%
 \nn 
 && \hspace*{-1.5cm} +\,
 4 g^2 \CA \CF  
 \Tint{ \{PQ\} }
  \biggl\{
      (D-2)
      \biggl[
       \frac{1}{P^4}
       - \frac{K^2 + 4 p^{ }_n ( p^{ }_n - k^{ }_n ) }{P^4(P-K)^2}
      \biggr]
      \biggl[ 
       \frac{1}{Q^2} - \frac{1}{(Q-P)^2} 
      \biggr]
%%%%%%%%%%%%%%%%%%%%%
  \nn & & \;
  - \, \frac{D-4}{Q^2(Q-P)^2}
    \biggl[
     \frac{1}{P^2} - \frac{1}{ (P-K)^2}  
    \biggr]
  - \frac{\fr12\! (D-6) (K^2 - k_n^2)}{P^2(P-K)^2Q^2(Q-K)^2}
%%%%%%%%%%%%%%%%%%%%%
  \nn[2mm] & & \; + \, 
  \frac{(D-6) K^2 - 2(D-2)K\cdot Q - 4 (D-4) p^{ }_n k^{ }_n
                                   + 4 (D-2) q^{ }_n k^{ }_n }
  {P^2 (P-K)^2 Q^2 (Q-P)^2}
%%%%%%%%%%%%%%%%%%%%%
  \nn[2mm] & & \; + \, 
  \frac{K^4 - 2 K^2 k_n^2 - 2 (D-4) K^2 p^{ }_n q^{ }_n + 2 (D-2) K^2 p_n^2}
  {P^2 (P-K)^2 Q^2 (Q-K)^2 (Q-P)^2 }
  \biggr\} \la{rho00}
 \;. 
\ea
The spectral functions corresponding to all 
structures here are worked out in ref.~\cite{gj}.

%%%%%%%%%%%%%%%%%%%%%%%%%%%% SECTION %%%%%%%%%%%%%%%%%%%%%%%%%%%%%%%%%%%%
%
\section{Theoretical considerations}
\la{se:limits}

%%%%%%%%%%%%%%%%%%%%%%%%%%%% SUBSECTION %%%%%%%%%%%%%%%%%%%%%%%%%%%%%%%%%
%
\subsection{OPE limit}
\la{ss:reduce}

We now take an imaginary part of
\eqs\nr{rhoV} and \nr{rho00} 
according to \eq\nr{cut}. Analytic results
can be obtained by considering $|\ko\pm k| \gg \pi T$~\cite{simon}. 
Limiting values for 
the ``master'' structures in \eq\nr{rhoV}
were given in appendix~B of ref.~\cite{relativistic}. 
The additional ones appearing in \eq\nr{rho00} can be determined by  
making use of techniques described in ref.~\cite{bulk_ope},
and are listed in ref.~\cite{gj}.

Inserting the expansions, we find that all 
$1/\epsilon$-divergences, the corresponding logarithms, 
as well as thermal corrections proportional to 
$
 \int_p \frac{\nB^{ }}{16\pi p}
$ 
or 
$
 \int_p \frac{\nF^{ }}{16\pi p}
$, 
cancel
($\nB^{ }$ and $\nF^{ }$ are the Bose and Fermi distributions, 
respectively). 
The remainders read
\ba
 \rhoV^{ } & = &  \frac{ \Nc^{ }M^2 }{4\pi}  
 + 4 g^2 \CF \Nc 
 \biggl\{ 
  \frac{3M^2}{4(4\pi)^3}  
  +  \int_p \frac{p}{\pi}
     \frac{(4 \nF^{ } - \nB^{ })(\ko^2 + \frac{k^2}{3})}{3M^4}
 \biggr\}
 + \rmO\biggl( \frac{T^6}{M^4} \biggr)
 \;, \hspace*{5mm} \\ 
 \rho^{ }_{00} & = &  - \frac{ \Nc^{ } k^2 }{12\pi} 
 - 4 g^2 \CF \Nc 
 \biggl\{ 
  \frac{k^2}{4(4\pi)^3}  
  +  \int_p \frac{p}{\pi}
     \frac{(4 \nF^{ } - \nB^{ }) k^2}{9M^4}
 \biggr\}
 + \rmO\biggl( \frac{T^6}{M^4} \biggr)
 \;. 
\ea
Thereby, in accordance with the general argument
in ref.~\cite{harvey}, the combination in \eq\nr{harvey}
displays only a thermal correction: 
\be
 \rhoH^{ }  =  
  4 g^2 \CF \Nc 
    \int_p \frac{p}{\pi}
     \frac{4(4 \nF^{ } - \nB^{ })k^2}{9M^4}
 + \rmO\biggl( \frac{T^6}{M^4} \biggr)
 \;. \la{rhoH_asymp}
\ee
The integrals evaluate to 
$
 \int_p \frac{p\,\nB^{ }}{\pi} = \frac{ \pi T^4}{30}
$
and 
$
 \int_p \frac{p\,\nF^{ }}{\pi} = \frac{ 7 \pi T^4}{240}
$, 
so that $\rhoH^{ }$ approaches zero from the positive side.
We note, however, that the Operator Product Expansion (OPE)
shows poor convergence; 
the actual $\rhoH^{ }$ switches from negative to positive
only around $\omega \sim 20 T$.

%%%%%%%%%%%%%%%%%%%%%%%%%%%% SUBSECTION %%%%%%%%%%%%%%%%%%%%%%%%%%%%%%%%%
%
\subsection{LPM limit}
\la{ss:lpm_full}

We next consider an ``opposite'' limit to that in 
\se\ref{ss:reduce}, namely $M^2 \to 0^\pm_{ }$. The spatial
momentum is kept fixed, with a value $k\sim \pi T$. In this limit
the spectral function needs to be resummed in order to account for
the Landau-Pomeranchuk-Migdal (LPM) effect. 

Close to the light cone,  
it is often convenient to represent the two polarizations 
in a basis different from that in \eq\nr{harvey}. 
Specifically, we define the ``transverse'' and ``longitudinal'' 
spectral functions as 
\be
 \rhoT^{ } \; \equiv \;
 \sum_{i \perp \vec{k}} \rho^{ }_{ii} 
 \;, \quad
 \rhoL^{ } \; \equiv \; 
 \rho^{ }_{\parallel} + \rho^{ }_{00}
 \;, 
 \la{rhoTL}
\ee
where $\perp$ and $\parallel$ refer to the components perpendicular
and parallel to $\vec{k}$. 
Current conservation implies that 
$ 
 \rhoL^{ } = - (M^2/k^2) \rho^{ }_{00}
$, 
and in this basis \eq\nr{harvey} becomes
\be
 \rhoV^{ } \; = \; \rhoT^{ } + \rhoL^{ }
 \;, \quad
 \rhoH^{ } = \rhoT^{ } - (D-2) \rhoL^{ }
 \;. \la{rhoH}
\ee 

Following ref.~\cite{agmz}, the LPM-resummed spectral functions
$\rho^{ }_i$, with $i=T,L$, read
\ba
% && \hspace*{-1.5cm}
 \left. \rho^{ }_i \right|^\rmi{full}_\rmii{LPM} 
 & \equiv &  
 - \frac{\Nc}{\pi}
 \int_{-\infty}^{\infty} \! {\rm d}\epsilon \, 
 \bigl[ 1-\nF^{ }(\epsilon)-\nF^{ }(\ko-\epsilon) \bigr]
 \nn 
 & \times &  
 \lim_{\vec{y}\to \vec{0}}  \mathbbm{P} 
 \biggl\{ 
   \frac{M^2
   \delta^{ }_{i,\rmii{$L$}}
   \im [g(\vec{y})]
   }{\ko^2}
  + 
   \frac{[ \ko^2 - 2 \epsilon(\ko^{ } - \epsilon) ] 
   \delta^{ }_{i,\rmii{$T$}}
   \im [\nabla^{ }_{\perp}\cdot \vec{f}(\vec{y})] }
        {2 \epsilon^2 (\ko^{ } - \epsilon)^2}  
 \biggr\} 
 \;, \la{final_from_LPM} \hspace*{5mm}
\ea
where $\mathbbm{P}$ stands for a principal value, and 
$g$ and $\vec{f}$ are Green's functions satisfying
\be
 \bigl( \hat{H} + i 0^+_{ }\bigr) g(\vec{y}) \; = \; \delta^{(2)}(\vec{y})
 \;, \quad
 \bigl( \hat{H} + i 0^+_{ }\bigr) \vec{f}(\vec{y}) 
 \; = \; -\nabla^{ }_{\perp} \delta^{(2)}(\vec{y}) 
 \;. 
\ee
The operator $\hat{H}$ acts in the plane transverse to light-like propagation, 
\be
 \hat{H} = -\frac{M^2}{2\ko^{ }}
 + \frac{\ko^{ }
  ( m_\infty^2 - \nabla_\perp^2 )
  }{2\epsilon(\ko^{ }- \epsilon)}
 + i \gE^2 \CF^{ }
   \int\! \frac{{\rm d}^2 \vec{q}}{(2\pi)^2}
  \bigl( 1 - e^{i \vec{q}\cdot\vec{y}}\bigr)
  \biggl( 
   \frac{1}{q^2} - \frac{1}{q^2 + \mE^2}
  \biggr)
 \;, \la{hatH}
\ee
where $ m_\infty^2 $ is an ``asymptotic'' quark thermal mass, 
given in \eq\nr{minfty}, whereas $\gE^2 \simeq g^2 T$ and 
$\mE^2 \simeq g^2 T^2 (\frac{\Nc^{ } }{ 3 } + \frac{ \Nf^{ } }{6} )$
are parameters of a dimensionally reduced 
effective theory~\cite{dr1,dr2,generic}.

%%%%%%%%%%%%%%%%%%%%%%%%%%%% SUBSECTION %%%%%%%%%%%%%%%%%%%%%%%%%%%%%%%%%
%
\subsection{Prediction for IR-singularities around the light cone}
\la{ss:lpm_expanded}

An interesting application of 
\eqs\nr{final_from_LPM}--\nr{hatH} is that 
by re-expanding them as a power series in $g^2$, we can find
out what kind of singularities the strict 2-loop results~\cite{gj} should 
contain close to the light cone. For this purpose, we follow
a procedure described 
in \se{5.1} of ref.~\cite{interpolation}. 
At zeroth order in $g$,  
the expressions become
\be
 \rhoT^{ } \big|_\rmii{LPM}^{(g^0)}
 \; = \; 
 \frac{\Nc M^2}{4\pi \ko^3} \, 
 \bigl( 
 \mathcal{I}^{ }_1 - 
 \mathcal{I}^{ }_2 
 \bigr)
 \;, \qquad
 \rhoL^{ } \big|_\rmii{LPM}^{(g^0)} 
 \; = \; 
 \frac{\Nc M^2}{4\pi \ko^3} \, \mathcal{I}^{ }_2 
 \;, \la{LPM_LO}
 \ee
where 
\ba 
 \mathcal{I}^{ }_1 & \equiv & 
 \biggl\{ 
   \theta(M^2) \int_0^{ \ko^{ }} \! {\rm d}\epsilon 
 - \theta(-M^2)
 \biggl[  \int_{-\infty}^0 + \int_{\ko^{ }}^{\infty} \biggr] 
 \, {\rm d}\epsilon
 \biggr\} 
 \bigl[ \nF^{ }(\epsilon - \ko^{ }) - \nF^{ }(\epsilon) \bigr]
 \, \ko^2 
 \nn 
 & = & 
 \theta(M^2) \, 
 \ko^3
 + 2 \ko^2 T \bigl[ \lnf(\ko^{ }) - \lnf(0) \bigr]
 \;, \la{I_1} \\ 
%%%%%%%%%%%%%%%%%%%
 \mathcal{I}^{ }_2 & \equiv & 
 \biggl\{ 
   \theta(M^2) \int_0^{ \ko^{ }} \! {\rm d}\epsilon 
 - \theta(-M^2)
 \biggl[  \int_{-\infty}^0 + \int_{\ko^{ }}^{\infty} \biggr] \, {\rm d}\epsilon
 \biggr\} 
 \bigl[\nF^{ }(\epsilon - \ko^{ })  - \nF^{ }(\epsilon) \bigr]
 \, 2 \epsilon (\ko^{ }- \epsilon)
 \nn 
 & = & 
 \theta(M^2) \, 
 \frac{\ko^3}{3} 
 + 4 \ko^{ } T^2 
 \bigl[ \lif(\ko^{ }) + \sign(M^2)\, \lif(0) \bigr]
 + 8 T^3 
 \bigl[ \ltf(\ko^{ }) - \ltf(0) \bigr]
 \;. \la{I_2}
\ea
The polylogarithms appearing here were defined in \eq\nr{polylogs}.
Even though $\mathcal{I}^{ }_{1,2}$ are not analytic around the
light cone, \eq\nr{LPM_LO} vanishes there.

Given that the last term in \eq\nr{hatH} is of $\rmO(g^4)$, 
the corrections of $\rmO(g^2)$ are proportional to the 
parameter $m_\infty^2$. 
For $\rhoL^{ }$, we find no such correction: 
\be
 \rhoL^{ } \big|^{(g^2)}_\rmii{LPM} = 0 
 \;. \la{LPM_NLO_L}
\ee
For $\rhoT^{ }$, a correction is found which contains
a well-known logarithmic divergence as well as a finite part which is 
discontinuous across the light cone: 
\ba
% && \hspace*{-1cm} 
 \rhoT^{ } \big|^{(g^2)}_\rmii{LPM} 
 & = & 
 \frac{\Nc^{ } m_\infty^2 }{2\pi}
 \Biggl\{ 
   \biggl[ \frac{1}{2} - \nF^{ }(\ko^{ }) \biggr]
   \biggl( \ln\biggl| \frac{m_\infty^2}{M^2}  \biggr| - 1 \biggr)
 \nn 
 & + &  
 \biggl[
 \theta(M^2) \, 
 \int_0^{\ko^{ }}\! {\rm d}\epsilon  
 - 
 \theta(-M^2) 
 \biggl( \int_{-\infty}^{0} + \int_{\ko^{ }}^{\infty} \biggr)
 \, {\rm d}\epsilon  
 \biggl]
 \nn 
 & \times &  
 \biggl[ 
 \frac{
   \nF^{ }(\epsilon) -  
   \nF^{ }(0) + 
   \nF^{ }(\ko^{ }-\epsilon) -  
   \nF^{ }(\ko^{ })   
 }{\epsilon}
 + 
 \frac{ 
   \nF^{ }(\epsilon - \ko^{ }) -  
   \nF^{ }(\epsilon)
 }{\ko^{ }}
 \biggr] \Biggr\}  
 \;. \la{LPM_NLO_T}
\ea
The integral on the last row is defined in the sense of a principal
value at large $|\epsilon|$, where terms $\sim 1/\epsilon$
cancel due to contributions from negative and positive $\epsilon$. 
Eq.~\nr{LPM_NLO_T} predicts that the strict 2-loop spectral function 
is discontinuous across the light cone, specifically
\be
 \Bigl\{  
  \lim_{\ko^{ }\to k^+_{ }} - \lim_{\ko^{ }\to k^-_{ }}
 \Bigr\} 
 \rhoT^{ } \big|^{(g^2)}_\rmii{ }
 \; = \; 
 \frac{g^2 T^2 \Nc^{ } \CF^{ }}{8\pi}
 \int_{-\infty}^\infty \!
 {\rm d}\epsilon 
 \, 
 \mathbbm{P} \biggl\{  
 \bigl[  \nF^{ }(\epsilon - k) - \nF^{ }(\epsilon) \bigr]
 \biggl( \frac{1}{k} - \frac{1}{\epsilon} \biggr) \biggr\} 
 \;, \la{disc_prediction}
\ee
where we inserted the definition of $m_\infty^2$ from \eq\nr{minfty}.

%%%%%%%%%%%%%%%%%%%%%%%%%%%% SUBSECTION %%%%%%%%%%%%%%%%%%%%%%%%%%%%%%%%%
%
\subsection{Matching of IR-singularities around the light cone}
\la{ss:matching}

It is a basic premise of LPM resummation that close to the light cone
it eliminates the IR singularities that plague the perturbative series. 
In other words, when \eq\nr{LPM_NLO_T} is subtracted from
the 2-loop expression, 
the remainder should be non-singular.\footnote{%
 The 2-loop expressions and their IR singularities
 can also be checked in the regime $\ko^{ }, k \ll \pi T$, 
 where they match the imaginary part of the photon HTL self-energy, 
 computed up to NLO in ref.~\cite{nlo_htl}.
 } 

The logarithmic singularities and
discontinuities originate from two structures,
both contained in \eq\nr{rhoV}. 
The first source are 
the factorized terms on the second line. Setting $D\to 4$
and identifying
\be
  m_\infty^2
  \; \equiv \; 
  g^2 \CF^{ } (D-2)\, \Tint{ \{Q\} }       \biggl[ 
     \frac{1}{(Q-P)^2} - \frac{1}{Q^2} 
      \biggr]
 \; \stackrel{D=4}{=} \; \frac{g^2 \CF^{ }T^2}{4} 
 \;, \la{minfty}
\ee
the discontinuity from the second line is 
\be
 \rhoV^{ } |^\rmi{disc}_{ } \supset 8 \Nc^{ }m_\infty^2 
 \im \biggl[ 
 \Tint{ \{P\} } \frac{1}{P^2(P-K)^2}
 \biggr]^{ }_{k^{ }_n \to -i [\ko^{ }+ i 0^+]}
 \;. \la{disc_fz}
\ee
Carrying out the Matsubara sum and taking the cut, we find
\ba
 && \hspace*{-1.3cm}
 \im \biggl[ 
 \Tint{ \{P\} } \frac{1}{P^2(P-K)^2}
 \biggr]^{ }_{k^{ }_n \to -i [\ko^{ }+ i 0^+]}
 \nn 
 & = & 
 \frac{1}{16\pi k}
 \biggl\{ 
 \theta(M^2)
 \int_{\km^{ }}^{\kp^{ }} 
 \!  {\rm d}\epsilon
 \; - \; 
 \theta(-M^2)
 \biggl[\int_{-\infty}^{\km^{ }} + \int_{\kp^{ }}^{\infty} \biggr]
 \, {\rm d}\epsilon
 \biggr\} 
 \, 
 \bigl[ \nF^{ }(\epsilon - \ko^{ }) - \nF^{ }(\epsilon) \bigr]
 \;. \la{Ib}
\ea
The discontinuity
of this expression precisely matches the terms $\propto 1/k$
in \eq\nr{disc_prediction}. 

The other terms of \eq\nr{disc_prediction} match the 
spectral function denoted by
\be
 \rho^{ }_{\mathcal{I}_\rmii{h'}}
 \; \equiv \; 
 \im\biggl[ 
  \Tint{ \{PQ\} } \frac{2 K\cdot Q}{P^2(P-K)^2Q^2(Q-P)^2}
 \biggr]^{ }_{k^{ }_n \to -i [\ko^{ }+ i 0^+]}
 \;, 
\ee
which in ref.~\cite{nlo} was shown to reproduce the logarithmic
singularity shown on the first row of \eq\nr{LPM_NLO_T}. 
Here we focus on the discontinuity. The expression obtained after
carrying out the Matsubara sums is given in 
\eq(B.84) of ref.~\cite{relativistic}, with 
$\sigma^{ }_1 = \sigma^{ }_2 = \sigma^{ }_{4} = -$,  $\sigma^{ }_5 = +$.

The discontinuity comes from the ``virtual'' part of 
$
 \rho^{ }_{\mathcal{I}_\rmii{h'}}
$
(the last lines of \eq(B.84)). 
If we define 
\ba
 \Phi(\ko^{ },\epsilon^{ }_p,\vec{k}\cdot\vec{p})
 & \equiv & 
 \int_\vec{q} 
 \biggl[
   \frac{1 - \nF^{ }(\epsilon^{ }_q) + \nB^{ }(E^{ }_{qp})}
        {4\epsilon^{ }_q E^{ }_{qp}}
   \biggl(
     \frac{\ko^{ }\epsilon^{ }_q + \vec{k}\cdot\vec{q}}
     {\epsilon^{ }_p + \epsilon^{ }_q + E^{ }_{qp}}
  + 
     \frac{- \ko^{ }\epsilon^{ }_q + \vec{k}\cdot\vec{q}}
     {-\epsilon^{ }_p + \epsilon^{ }_q + E^{ }_{qp}}
   \biggr) 
 \nn 
 & & \quad + \,  
   \frac{\nF^{ }(\epsilon^{ }_q) + \nB^{ }(E^{ }_{qp})}
        {4\epsilon^{ }_q E^{ }_{qp}}
   \biggl(
     \frac{\ko^{ }\epsilon^{ }_q - \vec{k}\cdot\vec{q}}
     {\epsilon^{ }_p - \epsilon^{ }_q + E^{ }_{qp}}
  + 
     \frac{\ko^{ }\epsilon^{ }_q + \vec{k}\cdot\vec{q}}
     {\epsilon^{ }_p + \epsilon^{ }_q - E^{ }_{qp}}
   \biggr) 
 \biggr]
 \;, \la{Phi}
\ea
and denote for brevity $\delta^{ }_x \equiv \delta(x)$, 
the virtual part reads 
\ba
 && \hspace*{-0.5cm} 
 \rho^\rmi{(v)}_{\mathcal{I}_\rmii{h'}} =
 \int_\vec{p} \frac{2\pi}{ 4 \epsilon^{ }_p \epsilon^{ }_{pk}}
 \la{Ihp_v} \\ 
 & 
 \times 
 & \!\! \Bigl\{ \Phi(\ko^{ },\epsilon^{ }_p,\vec{k}\cdot\vec{p})
  \Bigl[ 
  \delta^{ }_{ \ko^{ } - \epsilon^{ }_p - \epsilon^{ }_{pk} } 
  \bigl[ 1 - \nF^{ }(\epsilon^{ }_p) - \nF^{ }(\epsilon^{ }_{pk})\bigr]
 + 
  \delta^{ }_{ \ko^{ } - \epsilon^{ }_p + \epsilon^{ }_{pk} }
  \bigl[ \nF^{ }(\epsilon^{ }_p) - \nF^{ }(\epsilon^{ }_{pk})\bigr]
  \Bigr]
 \nn 
%%%%%%%%%%%%%%%%
 & - & 
 \Phi(-\ko^{ },\epsilon^{ }_p,\vec{k}\cdot\vec{p})
 \Bigl[
  \delta^{ }_{ \ko^{ } + \epsilon^{ }_p + \epsilon^{ }_{pk} } 
  \bigl[ 1 - \nF^{ }(\epsilon^{ }_p) - \nF^{ }(\epsilon^{ }_{pk})\bigr]
 + 
  \delta^{ }_{ \ko^{ } + \epsilon^{ }_p - \epsilon^{ }_{pk} } 
  \bigl[ \nF^{ }(\epsilon^{ }_p) - \nF^{ }(\epsilon^{ }_{pk})\bigr]
 \Bigr] \Bigr\}
 \;. \nonumber 
\ea

Now, the $\delta$ constraints 
in \eq\nr{Ihp_v} are equivalent to those emerging from 
\eq\nr{disc_fz}. Recalling 
$\epsilon^{ }_{pk} \equiv |\vec{p-k}|$,
a key observation is that if we approach the 
light cone from above ($\ko^{ }\to k^+$), only the first channel
contributes, and the contribution emerges from the domain
$\epsilon^{ }_{pk} \approx k - \epsilon^{ }_p $, i.e.\ 
$\vec{p} \parallel \vec{k}$ and $\epsilon^{ }_p < k$. If we approach the 
light cone from below ($\ko^{ }\to k^-$), there is a contribution
from the second channel, which emerges from the domain
$\epsilon^{ }_{pk} \approx \epsilon^{ }_p - k$, i.e.\ 
$\vec{p} \parallel \vec{k}$ and $\epsilon^{ }_p > k$. 
Below the light cone there is also a contribution
from the fourth channel, but now it emerges from the domain
$\epsilon^{ }_{pk} \approx \epsilon^{ }_p + k$, i.e.\ 
$-\vec{p} \parallel \vec{k}$ and $\epsilon^{ }_p > 0$. 
In total we get 
\ba
 \Bigl\{  
  \lim_{\ko^{ }\to k^+_{ }} - \lim_{\ko^{ }\to k^-_{ }}
 \Bigr\} \rho^{\rmi{(v)}}_{\mathcal{I}_\rmii{h'}}
 & = & 
 \frac{1}{8\pi k} 
 \int_0^{\infty} \! {\rm d}\epsilon^{ }_p \, 
 \Bigl\{ 
   \bigl[ \nF^{ }(\epsilon^{ }_p - k) - \nF^{ }(\epsilon^{ }_p) \bigr]
   \, \Phi(k,\epsilon^{ }_p, k \epsilon^{ }_p)
 \nn 
 & & \hspace*{1cm} 
 + \, 
   \bigl[ \nF^{ }(\epsilon^{ }_p ) - \nF^{ }(\epsilon^{ }_p + k) \bigr]
   \, \Phi(-k,\epsilon^{ }_p, -k \epsilon^{ }_p) 
 \Bigr\}  
 \;. \la{disc}
\ea
Carrying out the angular integral in \eq\nr{Phi} and setting  
subsequently $\ko^{ }$ and $\vec{k}\cdot\vec{p}$ to the values 
required by \eq\nr{disc}, it can be verified that the UV-divergent
vacuum term and the IR-sensitive\footnote{% 
 It is practical to regularize IR divergences by setting 
 $E_{qp}^2 \equiv (\vec{q-p})^2 + \lambda^2$, 
 with $\lambda\to 0$ at the end. 
 }
thermal terms drop out. 
Moreover, the integral over $q$ yields
\be
 \Phi(k,\epsilon^{ }_p,k\epsilon^{ }_p)
 \;=\; 
 - \Phi(-k,\epsilon^{ }_p,-k\epsilon^{ }_p)
 \;=\; 
 \frac{k}{2\epsilon^{ }_p}
 \int_\vec{q} \frac{\nF^{ }(\epsilon^{ }_q) + \nB^{ }(\epsilon^{ }_q)}
                   {\epsilon^{ }_q}
 = \frac{k T^2}{16 \epsilon^{ }_p}
 \;. 
\ee
Going over to a variable $\epsilon = \pm \epsilon^{ }_p$
for convenience, we subsequently find 
\be
 \Bigl\{  
  \lim_{\ko^{ }\to k^+_{ }} - \lim_{\ko^{ }\to k^-_{ }}
 \Bigr\} \rho^{\rmi{(v)}}_{\mathcal{I}_\rmii{h'}}
 \; = \; 
 \frac{T^2}{128\pi}
 \int_{-\infty}^\infty \!
 {\rm d}\epsilon 
 \, 
 \mathbbm{P} \biggl\{ 
 \frac{ \nF^{ }(\epsilon - k) - \nF^{ }(\epsilon)
      }{\epsilon} \biggr\} 
 \;. \la{disc_Ihp}
\ee
Multiplying by  
$- 16 g^2 \Nc^{ }\CF^{ }$ from \eq\nr{rhoV}, 
the part $\propto -1/\epsilon$
of \eq\nr{disc_prediction} is reproduced. 

%%%%%%%%%%%%%%%%%%%%%%%%%%%% SUBSECTION %%%%%%%%%%%%%%%%%%%%%%%%%%%%%%%%%
%
\subsection{Sum rules}
\la{ss:sum}

A traditional further constraint on spectral functions is offered
by sum rules (cf.,\ e.g.,\ ref.~\cite{sumrule} and references therein). 
Unlike the OPE and LPM
limits, the sum rules are sensitive to the complete frequency domain. 
However, for $\rhoV^{ }$ they
are of limited value, as they require the subtraction of 
poorly known vacuum parts (containing a dense spectrum of resonances). 
In contrast, a nice and convergent sum rule 
can be obtained for $\rhoH^{ }$~\cite{harvey}: 
\be
 \int_0^\infty \! {\rm d}\omega \, \omega \,
 \rhoH^{ }(\omega,k) = 0 
 \;. \la{sum_rule}
\ee

We have used our perturbative results in order to test which frequency
domain gives a contribution to \eq\nr{sum_rule}. It must be 
noted that $\rhoH^{ }$ displays a highly non-trivial structure, changing
sign twice: $\rhoH^{ }$ is positive at $\omega \le k$, becomes
negative at $\omega \gsim k$ as is necessary for the cancellation required
by \eq\nr{sum_rule}, but then again becomes positive when 
$|\omega - k | \gg \pi T$, as shown by \eq\nr{rhoH_asymp}. 
While we have verified that the sum rule is satisfied within 
numerical uncertainties by our strict 2-loop result and can also be
imposed once resummations are included (cf.\ below), we also see
that the asymptotics plays an important role, with 
the domain $\omega \ge 20 T$ giving a substantial contribution to 
the absolute value of the integral. 

%%%%%%%%%%%%%%%%%%%%%%%%%%%% SECTION %%%%%%%%%%%%%%%%%%%%%%%%%%%%%%%%%
%
\section{Comparison with lattice data}
\la{se:lattice}

%%%%%%%%%%%%%%%%%%%%%%%%%%%% SUBSECTION %%%%%%%%%%%%%%%%%%%%%%%%%%%%%%%%%
%
\subsection{Summary: resummed spectral functions}

%%%%%%%%%%%%%%%%%%%%%%%%%%%%%%%%% FIGURE %%%%%%%%%%%%%%%%%%%%%%%%%%%%%%%%%
\begin{figure}[t]

\hspace*{-0.1cm}
\centerline{%
 \epsfysize=7.5cm\epsfbox{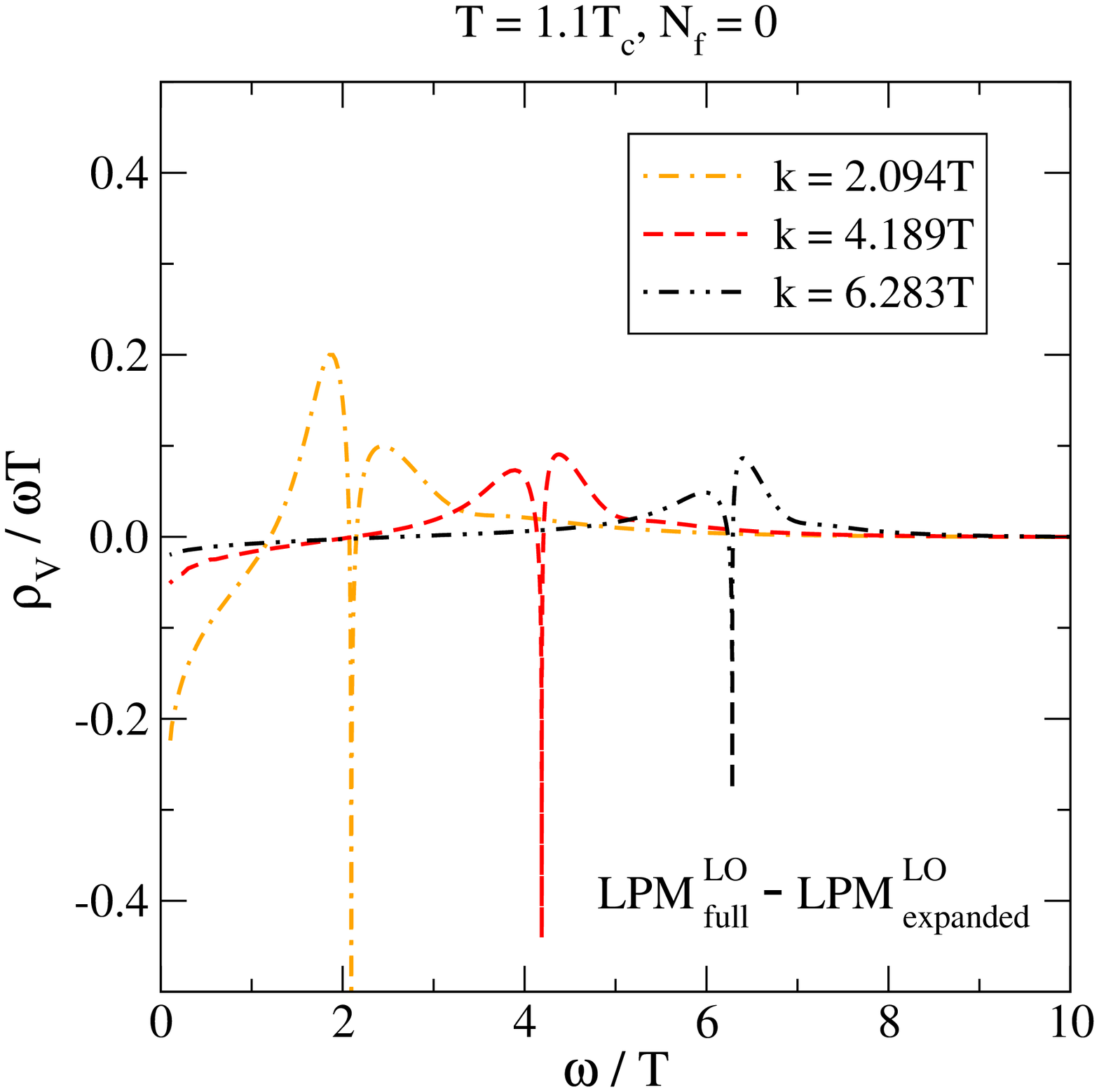}%
 \hspace{0.1cm}%
 \epsfysize=7.5cm\epsfbox{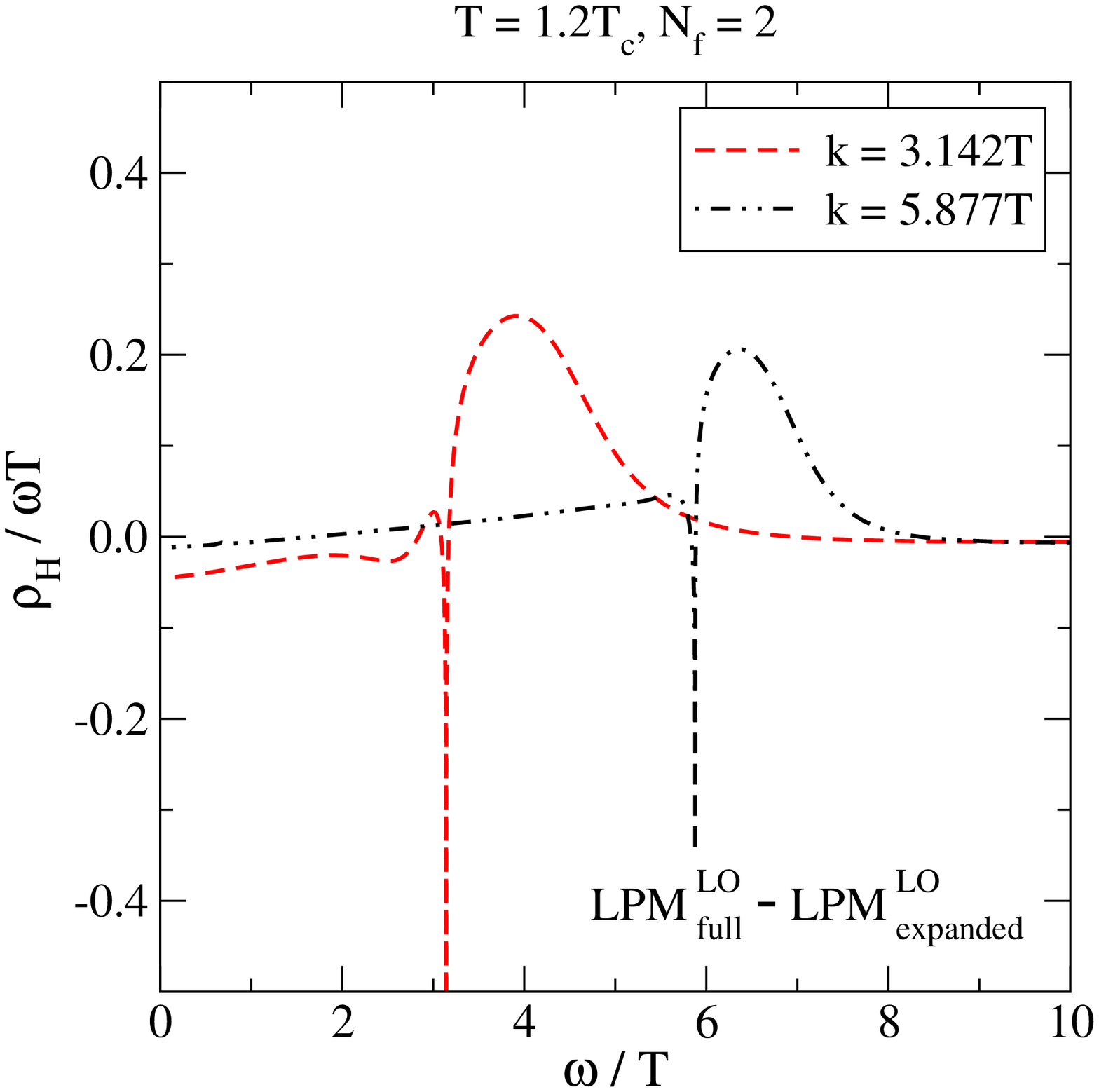}
}

\caption[a]{\small
 Left: 
 the modification of $\rhoV^{ }$ (cf.\ \eq\nr{harvey}) 
 by LPM-resummation (cf.\ \se\ref{ss:lpm_full}), 
 for $\bmu = \bmu^{ }_\rmi{opt}$. 
 The re-expanded version of the LPM result
 (cf.\ \se\ref{ss:lpm_expanded}) has been subtracted
 in order to avoid double-counting once the result is combined 
 with the full unresummed 2-loop expression, 
 cf.\ \eq\nr{resummation}. The logarithmic singularity
 cancels in this combination.  
 Right: the same for $\rhoH^{ }$ (cf.\ \eqs\nr{harvey}, \nr{rhoH}). 
}

\la{fig:rhoV_LPM}
\end{figure}
%%%%%%%%%%%%%%%%%%%%%%%%%%%%%%%%%%%%%%%%%%%%%%%%%%%%%%%%%%%%%%%%%%%%%%%%%%%

Having discussed various limits and crosschecks of the spectral functions, 
we are now ready to put together estimates
for phenomenological purposes. 
The full resummed spectral functions 
($i \in \{V,H,T,L\}$)
are defined as 
\be
 \rho^{ }_i |_\rmii{NLO}^\rmi{resummed}
 \; \equiv \; 
 \rho^{ }_i |_\rmi{2-loop}^\rmi{strict}
 + 
 \bigl( 
 \rho^{ }_i |_\rmii{LPM}^\rmi{full}
 - 
 \rho^{ }_i |_\rmii{LPM}^\rmi{expanded}
 \, \bigr) 
 \times \phi
 \;, \la{resummation} 
\ee
where 
$ \rho^{ }_i |_\rmi{2-loop}^\rmi{strict} $
is from ref.~\cite{gj}; 
$ \rho^{ }_i |_\rmii{LPM}^\rmi{full} $
is from \se\ref{ss:lpm_full}; 
and 
$ \rho^{ }_i |_\rmii{LPM}^\rmi{expanded} 
 \equiv
 \rho^{ }_i |_\rmii{LPM}^{(g^0)}
 + 
 \rho^{ }_i |_\rmii{LPM}^{(g^2)}
$
is from \se\ref{ss:lpm_expanded}.
The function $\phi$, which should be unity if resummations were implemented
``exactly'', and must in any case equal unity in the IR domain, 
can be used to correct for the fact that 
kinematic simplifications pertinent only to the IR domain have
been employed in order to implement the resummation. 
Outside of this domain, we can use $\phi$ to switch off the 
resummation more rapidly than it would switch off otherwise. 
We find it practical to define 
$
 \phi^\rmii{LO}_{ } \equiv \theta(\omega^*_{ } - \omega)
$, 
where $\omega^*_{ }$ is chosen so that the second structure of 
\eq\nr{resummation} satisfies \eq\nr{sum_rule} (just like the first
structure does). The superscript LO stands for leading-order LPM
resummation, as described in \ses\ref{ss:lpm_full} and 
\ref{ss:lpm_expanded}, and we find that numerically
$ \omega^*_{ } \sim 15 ... 25 T$, depending on $k$. 
We also incorporate NLO LPM-resummed results from 
ref.~\cite{nlo2}, however for these the ``expanded'' version is not
available, and we thus impose a faster cutoff away from the light cone, 
inspired by discussions in ref.~\cite{nlo2}, 
\be
 \phi^\rmii{NLO}_{ } \; \equiv \; 
 \theta(k - \omega) \, \frac{e^{\,\omega / T} -1 }{e^{\,k/T} - 1}
  + 
 \theta(\omega - k) \, \frac{ e^{\,k / T} }{e^{\,\omega/T}}
 \;. \la{phi} 
\ee

In order to display the practical effect of the resummation, 
consider the difference
$
 \rho^{ }_i |_\rmii{LPM}^\rmi{full}
 - 
 \rho^{ }_i |_\rmii{LPM}^\rmi{expanded}
$
at leading order. 
Results are shown in \fig\ref{fig:rhoV_LPM}. 
Prominent features are a logarithmic divergence
around light cone, cancelling the one from
$ \rho^{ }_i |_\rmi{2-loop}^\rmi{strict} $,  
as well as the vanishing of the correction  
when $\ko^{ }\to 0$ or $\ko^{ }\to \infty$
(in \fig\ref{fig:rhoV_LPM} the spectral function 
is divided by $\ko^{ }$).

A practical evaluation of the spectral function necessitates a choice
of the renormalization scale for the gauge coupling. 
Motivated by the arguments in ref.~\cite{qhat}, we may expect that
the physics of the IR domain is represented by a dimensionally reduced
description, whereby a fastest apparent convergence criterion
suggests~\cite{muT1,muT2} 
\be
 \bmu_\rmi{opt}^\rmi{($\Nf\! = \! 0$)} = 6.74 T
 \;,\quad
 \bmu_\rmi{opt}^\rmi{($\Nf\! = \! 2$)} = 8.11 T
% \;,\quad
% \bmu_\rmi{opt}^\rmi{($\Nf\! = \!3$)} = 9.08 T
 \;. \la{mu_opt_DR}
\ee
Away from the IR domain, the scale should be set by virtuality. 
In order to smoothly interpolate between these two possibilities, 
we choose 
\be
 \bmu^{ }_\rmi{opt} \; \equiv \; \sqrt{(\xi \pi T)^2 + |M^2|}
 \;, \la{mu_opt}
\ee
taking $\xi = 1$ for $\Nf = 0$ and
a larger $\xi = 2$ for $\Nf = 2$.
As these are on the low side compared with
\eq\nr{mu_opt_DR}, we vary $\bmu$ in the range 
$(1.0 ... 2.0)\times \bmu^{ }_\rmi{opt}$, noting that the gauge coupling
grows uncontrollably large for
$\bmu = 0.5 \bmu^{ }_\rmi{opt}$ ($\alphas^{ } > 0.5$). 
The gauge coupling is solved for from 
5-loop evolution~\cite{alph1,alph2,alph3}. 
We have verified that the results are stable if resorting to lower-order
running or modifying the interpolation 
in \eq\nr{mu_opt} while keeping the limits
at $\pi T \ll |M|$ and $\pi T \gg |M|$ fixed. 

At very large $\omega$, 
we let $\rhoV^{ }$ continuously cross into vacuum-like N$^{4}_{ }$LO 
perturbative behaviour~\cite{cond}. 
Such results can be inserted into \eq\nr{relation}, in order to 
construct $G^{ }_\rmii{$V$}$. 
For $\rhoH^{ }$ the vacuum tail is absent, nevertheless
the results for  $G^{ }_\rmii{$H$}$
are quite sensitive to a broad frequency range 
$0 \le \omega \lsim 30 T$.

%%%%%%%%%%%%%%%%%%%%%%%%%%%% SUBSECTION %%%%%%%%%%%%%%%%%%%%%%%%%%%%%%%%%
%
\subsection{Comparison with lattice data for $\Nf = 0$~\cite{photon}}
\la{ss:quenched}

We start the lattice comparison with the data that were produced and
analyzed in ref.~\cite{photon}. The correlator measured was 
\be
 G_\rmii{$V$}^{ }(\tau,{k}) 
 \; \equiv \; 
 \int_{\vec{x}}  
  e^{- i \vec{k}\cdot\vec{x}}
 \Big\langle 
 {\textstyle\sum_{i=1}^{3}}
     V^{i}_{ } (\tau,\vec{x}) \, 
     V^{i}_{ } (0) 
 -
     V^{0}_{ } (\tau,\vec{x}) \, 
     V^{0}_{ } (0) 
 \Big\rangle^{ }_\rmi{c}
 \;, \la{G_V}
\ee
where $V^{\mu}_{ }$ is the (Minkowskian) vector current and 
$\langle ... \rangle^{ }_\rmi{c}$ stands for the connected
contractions. In the continuum limit this correlator diverges at small $\tau$
and is conveniently normalized to the free result
\be
 \frac{ G^{ }_{\rmi{norm},\rmii{$V$}} }{6 T^3}
 \; \equiv \;
  \pi (1-2\tau T) \frac{1 + \cos^2(2\pi\tau T)}{\sin^3(2\pi \tau T)}
 + 
 \frac{2 \cos(2\pi \tau T)}{\sin^2(2\pi \tau T)}
 \;. \la{GVnorm}
\ee
For scale setting, we 
use $\Tc^{ }/\Lambdamsbar \simeq 1.24$, which 
has $\sim 10\%$ uncertainty~\cite{Tc}. 

%%%%%%%%%%%%%%%%%%%%%%%%%%%%%%%%% FIGURE %%%%%%%%%%%%%%%%%%%%%%%%%%%%%%%%%
\begin{figure}[t]

\hspace*{-0.1cm}
\centerline{%
 \epsfysize=7.5cm\epsfbox{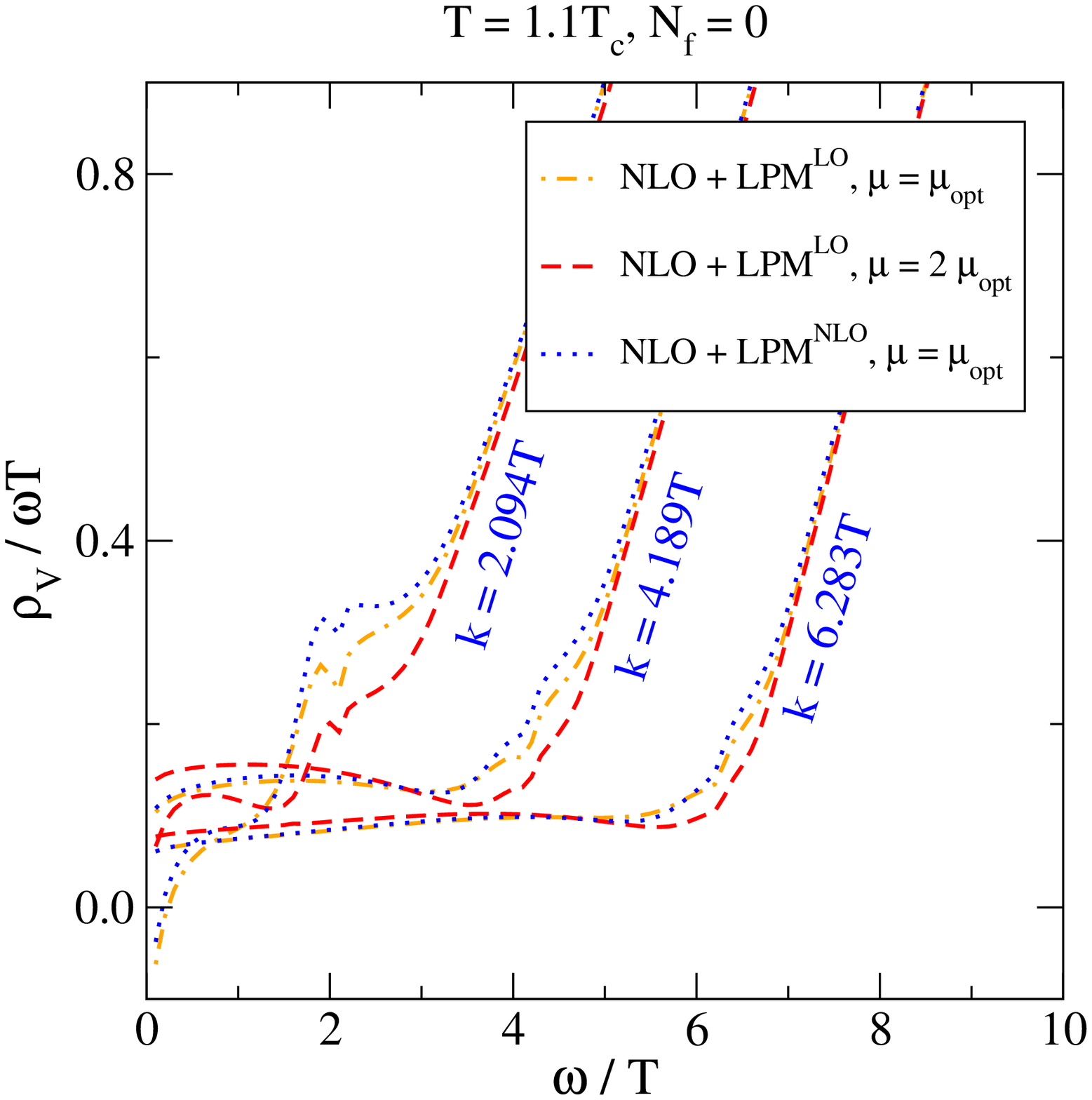}%
 \hspace{0.1cm}%
 \epsfysize=7.5cm\epsfbox{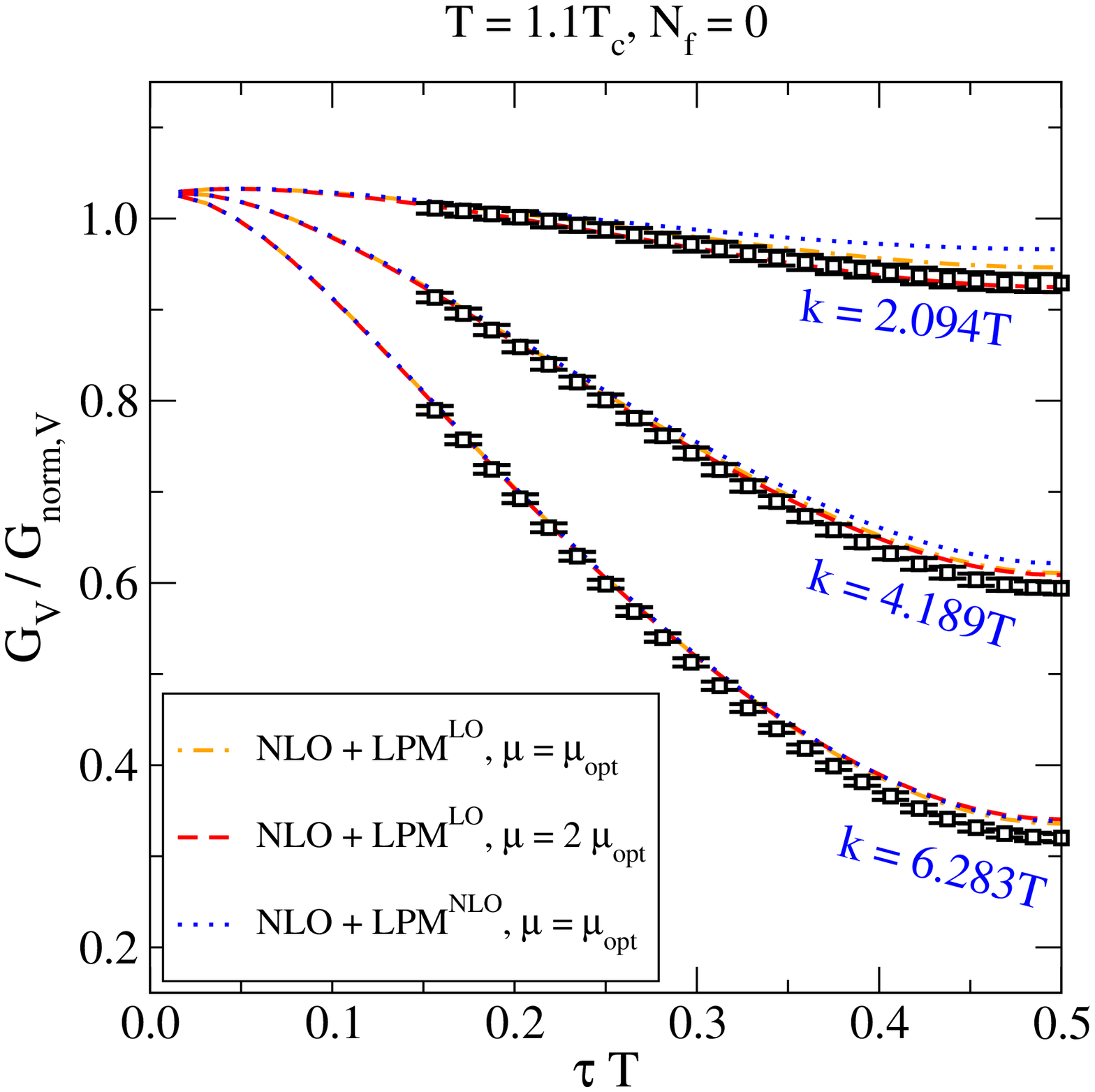}
}

\caption[a]{\small
 Results for $\rhoV^{ }$ (left) and $G^{ }_\rmii{$V$}$ (right)
 at $T = 1.1\Tc^{ }$ for $\Nf^{ }=0$, 
 the latter normalized to \eq\nr{GVnorm}.
 LPM$^\rmii{LO}_{ }$ refers to results 
 from 
 \ses\ref{ss:lpm_full}
 and 
 \ref{ss:lpm_expanded}, 
 employing the two scale choices
 $\bmu = \bmu^{ }_\rmi{opt}$  and 
 $\bmu = 2\bmu^{ }_\rmi{opt}$ (cf.\ \eq\nr{mu_opt}). 
 The notation LPM$^\rmii{NLO}_{ }$ indicates that 
 the contribution from ref.~\cite{nlo2} has been 
 added; in this case we use $\bmu = \bmu^{ }_\rmi{opt}$.
 The black squares are lattice results from ref.~\cite{photon}. 
 The spectral function can become negative at very small $\omega$ due to
 the subtraction of $\rho^{ }_{00}$ (cf.~\eq\nr{G_V}); the related physics 
 is discussed in more detail around \eq\nr{hydro}.  
}

\la{fig:GV_lattice_Nf0}
\end{figure}
%%%%%%%%%%%%%%%%%%%%%%%%%%%%%%%%%%%%%%%%%%%%%%%%%%%%%%%%%%%%%%%%%%%%%%%%%%%

Resummed NLO spectral functions $\rhoV^{ }$ are shown for three momenta
in \fig\ref{fig:GV_lattice_Nf0}(left), and the corresponding
imaginary-time correlators $G^{ }_\rmii{$V$}$ obtained from 
\eq\nr{relation} 
in \fig\ref{fig:GV_lattice_Nf0}(right), 
where they are also compared with lattice data. Despite the low
temperature, we observe a remarkable agreement.  
On close inspection, the perturbative curves are above the lattice ones, 
requiring a non-perturbative suppression of $\rhoV^{ }$. The same qualitative 
features persist at $T = 1.3\Tc^{ }$ (not shown), however the 
difference between the perturbative and lattice results is slightly 
smaller, as may be expected from a gradually decreasing $\alphas^{ }$. 
The conclusions that we draw from these observations are summarized
in \se\ref{se:concl}.

%%%%%%%%%%%%%%%%%%%%%%%%%%%% SUBSECTION %%%%%%%%%%%%%%%%%%%%%%%%%%%%%%%%%
%
\subsection{Comparison with lattice data for $\Nf^{ }=2$~\cite{harvey,new}}
\la{ss:unquenched}

Finally we move on to unquenched lattice data, obtained recently for 
$\Nf^{ } = 2$ in refs.~\cite{harvey,new}. In this case we concentrate
on the ultraviolet finite correlator
($
 \vec{k} \equiv k \vec{e}^{ }_z 
$) 
\be
 G_\rmii{$H$}^{ }(\tau,{k}) 
 \, \equiv \, 
 \int_{\vec{x}}  
  e^{- i k z}
 \Big\langle 
 {\textstyle\sum_{i=1}^{2}}
     V^{i}_{ } (\tau,\vec{x}) \, 
     V^{i}_{ } (0) 
  - 2 \Bigl[ 
     V^{z}_{ } (\tau,\vec{x}) \, 
     V^{z}_{ } (0) 
    - 
     V^{0}_{ } (\tau,\vec{x}) \, 
     V^{0}_{ } (0) 
     \Bigr]
 \Big\rangle^{ }_\rmi{c}
 \;. \la{G_H}
\ee
Let us stress again that the spectral functions corresponding to 
$G_\rmii{$V$}^{ }$ and $G_\rmii{$H$}^{ }$ agree on the light cone but
are substantially different away from it
(cf.\ \fig\ref{fig:GV_lattice_Nf0}(left) vs.\ 
\fig\ref{fig:GH_lattice_Nf2}(left)). 

%%%%%%%%%%%%%%%%%%%%%%%%%%%%%%%%% FIGURE %%%%%%%%%%%%%%%%%%%%%%%%%%%%%%%%%
\begin{figure}[t]

\hspace*{-0.1cm}
\centerline{%
 \epsfysize=7.5cm\epsfbox{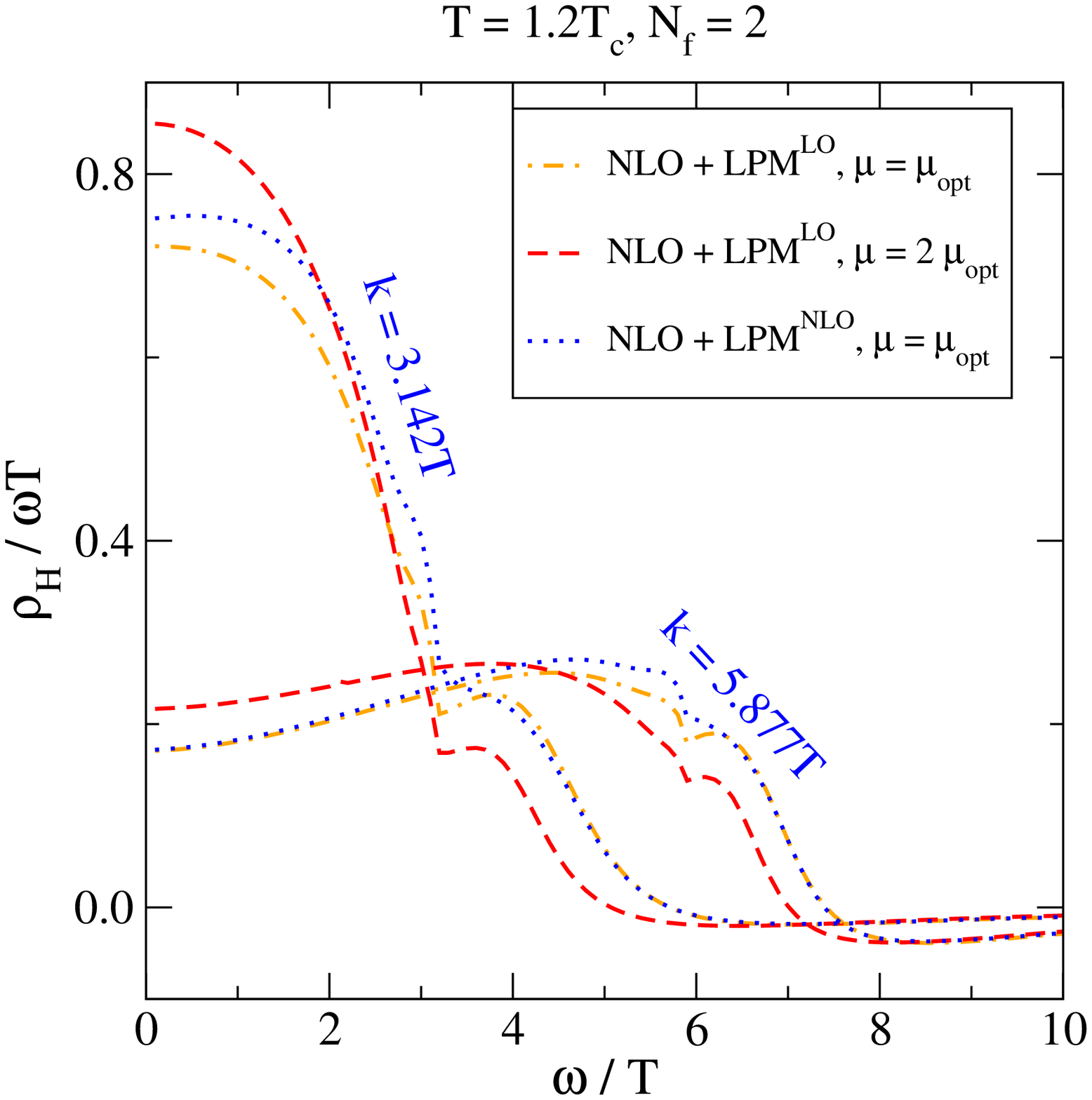}%
 \hspace{0.1cm}%
 \epsfysize=7.5cm\epsfbox{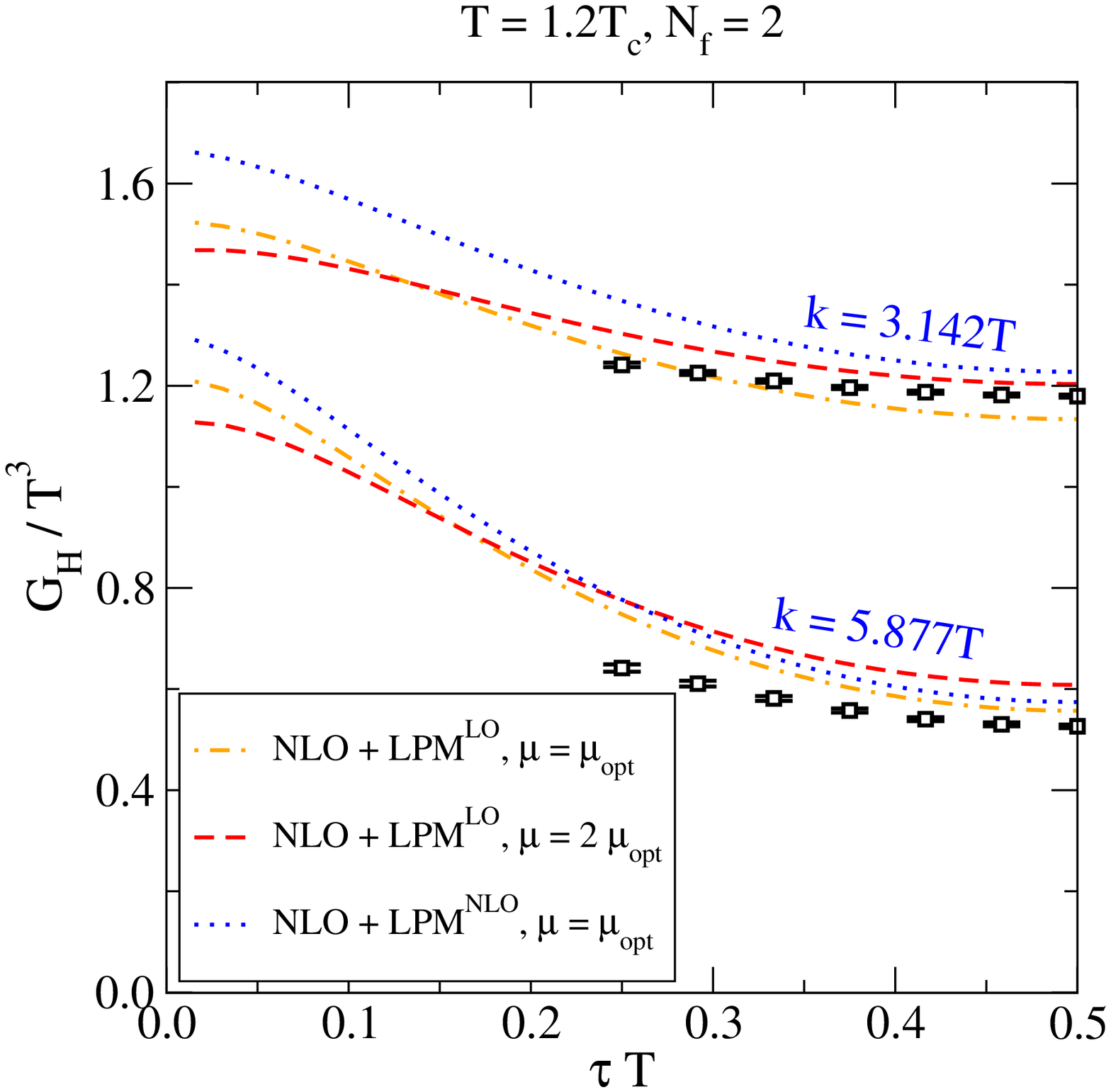}
}

\caption[a]{\small
 Results for $\rhoH^{ }$ (left) and $G^{ }_\rmii{$H$}$ (right)
 at $T \simeq 1.2\Tc^{ }$ for $\Nf^{ }=2$. The black squares
 are lattice data from refs.~\cite{harvey,new},
 multiplied by a factor $2 \chi /T^2$, 
 where $\chi \approx 0.88 T^2$~\cite{new}, in order to convert to our units.
 The notations LPM$^\rmii{LO}_{ }$ and LPM$^\rmii{NLO}_{ }$ and the
 scale choices are as in \fig\ref{fig:GV_lattice_Nf0}.
}

\la{fig:GH_lattice_Nf2}
\end{figure}
%%%%%%%%%%%%%%%%%%%%%%%%%%%%%%%%%%%%%%%%%%%%%%%%%%%%%%%%%%%%%%%%%%%%%%%%%%%

Again a comparison between perturbative and lattice results requires 
relating physical scales.
According to eq.~(3.1) of ref.~\cite{nf2_Tc},
$\Tc^{ } \simeq 167(25)$~MeV, with units set through 
$r^{ }_0 = 0.503(10)$~fm~\cite{nf2_r0}. 
Adopting a community average from ref.~\cite{nf2_Lambda},
{\it viz.} $r^{ }_0\Lambdamsbar \approx 0.75(10)$, 
yields
$
   \Tc^{ }/\Lambdamsbar \simeq 0.56
$
for $\Nf^{ }=2$, 
but with substantial $\sim 25\%$ uncertainties.
For the comparison, a susceptibility is needed
as well; we employ the recent continuum extrapolation
$\chi = 0.88(1) T^2$ from ref.~\cite{new}, consistent 
with classic expectations~\cite{susc}. 

The spectral function $\rhoH^{ }$ is shown in 
\fig\ref{fig:GH_lattice_Nf2}(left), 
and the corresponding imaginary-time correlator $G^{ }_\rmii{$H$}$
in \fig\ref{fig:GH_lattice_Nf2}(right). 
Like in \fig\ref{fig:GV_lattice_Nf0}(right), 
the lattice correlators fall in general below the perturbative curves. 
The uncertainties of the perturbative imaginary-time correlators,
as reflected by the scale dependence and the difference
between LPM$^\rmii{LO}_{ }$ and LPM$^\rmii{NLO}_{ }$ resummations,
are relatively speaking larger for $\Nf^{ }= 2$,
a manifestation of the fact that the dominant vacuum UV tail is absent 
and therefore the data is more sensitive to IR physics. 
Nevertheless it is comforting that the qualitative
pattern remains similar. 
The conclusions drawn from the comparison
are discussed in \se\ref{se:concl}.

%%%%%%%%%%%%%%%%%%%%%%%%%%%% SECTION %%%%%%%%%%%%%%%%%%%%%%%%%%%%%%%%%%%%
%
\section{Conclusions}
\la{se:concl}

Motivated by a comparison with lattice data,  
unresummed NLO (2-loop) vector spectral functions 
have recently been extended
into two new domains~\cite{gj}: 
below the light cone ($\omega < k$), 
and to a longitudinal polarization that vanishes
at the light cone but is non-zero elsewhere. 
Even if the spacelike domain, corresponding 
to deep inelastic scattering off a thermal medium, sounds academic, it is 
essential for a comparison with lattice data, given that imaginary-time
measurements get a large contribution from this region 
(cf.\ \eq\nr{relation}). The longitudinal polarization, in turn, is useful 
in the UV domain, as it permits to subtract the short-distance singularities 
from the lattice measurement (cf.\ \eq\nr{rhoH_asymp})~\cite{harvey}. 

With the 2-loop results at hand, 
they can be resummed close to the light cone
as specified in \eq\nr{resummation}
(parametrically, this is needed 
for $|\omega - k| \lsim \alphas^{ } T^2 / k$).  
Making use of methods developed in ref.~\cite{qhat}, this 
resummation has been worked out to NLO by now~\cite{nlo1,nlo2}, 
implying in this context corrections suppressed by 
$\sqrt{\alphas^{ }}$. We have
incorporated the latter corrections in our results, switching
them off away from the light cone when they lose their validity.

The comparison of the imaginary-time correlators following from 
the resummed NLO spectral functions against lattice data 
can be viewed as the inspection of 
many separate ``sum rules'', one for each $\tau$. 
Put together, this constrains the spectral
function in a non-trivial way. In particular, we find that
the correlators are affected by  
the choice of the renormalization scale of $\alphas^{ }$
(cf.\ \figs\ref{fig:GV_lattice_Nf0} and \ref{fig:GH_lattice_Nf2}). 
Reasonable agreement is obtained by scale choices reminiscent of
those originating from dimensional reduction (cf.\ \eq\nr{mu_opt_DR}). 

After fixing the renormalization scale, 
the perturbative results lie in general somewhat above the lattice data. 
Such a non-perturbative suppression confirms the previous finding 
based on a polynomial interpolation of $\rhoV^{ }$~\cite{photon}. 
At the same time the comparison of 
\figs\ref{fig:GV_lattice_Nf0}(right) and \ref{fig:GH_lattice_Nf2}(right)
testifies to the improved resolution power of 
the correlator $G^{ }_\rmii{$H$}$~\cite{harvey}, so we are looking forward to
final results from Pad\'e fits of $\rhoH^{ }$~\cite{new}.

It would be interesting to investigate if resummed NLO rates
embedded in hydrodynamical simulations of heavy ion collisions 
also overshoot the experimental results at small virtualities.  To our
knowledge this exercise has been implemented only on a rough level 
so far~\cite{hydro}, supporting however this
type of an overall trend. Nevertheless, it could still be that
the physical photon rate is well predicted or even underestimated
by the NLO result, if there is a large suppression of the 
spectral weigth in some other domain. The general expectation is that
strong interactions should suppress thermal fluctuations particularly
at small $\omega$ and $k$. 

We end by noting that $\rhoV^{ }$ of the spacelike domain has 
an interesting relation to 
the diffusion coefficient of hot QCD matter. 
For $\omega,k \ll T$, 
the general theory of statistical fluctuations
applies~\cite{landau}
and permits for 
a ``hydrodynamic'' prediction (cf.,\ e.g.,\ ref.~\cite{ht}), 
\be
 \frac{\rhoV^{ }(\omega,{k})}{\omega} 
 \quad \stackrel{\omega,k \ll T}{\approx} \quad
 \left( \frac{\omega^2 - k^2} {\omega^2 + D^2 k^4}
  + 2 \right) \chi^{ }_{ } D
 \;. \la{hydro}
\ee
Here $D$ is the diffusion 
coefficient and $\chi^{ }_{ }$ is a susceptibility, 
$\chi^{ }_{ } \equiv 
\int_0^\beta \! {\rm d}\tau \int_\vec{x}
\langle V^0(\tau,\vec{x}) V^0(0,\vec{0}) \rangle$.
It follows that the zero-frequency limit, 
$
 \lim_{\omega\to 0} {\rhoV^{ }(\omega,{k})} / {\omega}
$,
crosses zero at $k = 1/(\sqrt{2}D)$. 
The values extracted from our $\rhoV^{ } |_\rmii{NLO}^\rmi{resummed}$
this way are perfectly consistent with recent lattice estimates 
($DT \sim 0.2 ... 0.8$ at $T = 1.1\Tc$~\cite{Dlatt}) but differ from
strict LO perturbative determinations which incorporate further
resummations~\cite{Dpert}
($DT \approx 2.9$ at $T = 1.1\Tc$~\cite{photon}).\footnote{%
 We thank J.~Ghiglieri for pointing out that
 NLO corrections {\it \`a la} ref.~\cite{shear}
 do reduce the value of $DT$.
 } 

%%%%%%%%%%%%%%%%%%%%%%%%%%% SECTION %%%%%%%%%%%%%%%%%%%%%%%%%%%%%%%%%%
%
\section*{Acknowledgements}

We thank 
J.~Ghiglieri for discussions and for 
providing LPM$^\rmii{NLO}_{ }$ data from ref.~\cite{nlo2}, 
and 
M.~C\`e, T.~Harris, H.B.~Meyer, A.~Steinberg and A.~Toniato 
for providing lattice data from 
refs.~\cite{harvey,new}. This project was initiated
thanks to a communication by H.B.~Meyer, for which we express our gratitude.  
The work was partly supported by the Swiss National Science Foundation
(SNF) under grant 200020B-188712, and by the COST Action CA15213 THOR.

%%%%%%%%%%%%%%%%%%%%%%% APPENDIX %%%%%%%%%%%%%%%%%%%%%%%%%%%%%%%%%%%
%
\appendix
\renewcommand{\thesection}{Appendix~\Alph{section}}
\renewcommand{\thesubsection}{\Alph{section}.\arabic{subsection}}
\renewcommand{\theequation}{\Alph{section}.\arabic{equation}}

%%%%%%%%%%%%%%%%%%%%%%%%%%%%%%%%%%%%%%%%%%%%%%%%%%%%%%%%%%%%%%%%%%%%%%%%%%%


\begin{thebibliography}{99}

\bibitem{exp1}
  A.~Adare {\it et al.} [PHENIX Collaboration],
  {\it Dielectron production in Au$+$Au collisions 
  at $\sqrt{s_{\rm NN}} = 200$ GeV,}
  Phys.\ Rev.\ C {93} (2016) 014904
  [1509.04667].
  %%CITATION = doi:10.1103/PhysRevC.93.014904;%%

\bibitem{exp2}
  L.~Adamczyk {\it et al.} [STAR Collaboration],
  {\it Direct virtual photon production in Au+Au collisions 
  at $\sqrt{s_{\rm NN}}$ = 200 GeV,}
  Phys.\ Lett.\ B {770} (2017) 451
  [1607.01447].
  %%CITATION = doi:10.1016/j.physletb.2017.04.050;%%

\bibitem{exp3}
  J.~Adam {\it et al.} [ALICE Collaboration],
  {\it Direct photon production in Pb-Pb collisions 
  at $\sqrt{s_{\rm NN}} =$ 2.76 TeV,}
  Phys.\ Lett.\ B {754} (2016) 235
  [1509.07324].
  %%CITATION = doi:10.1016/j.physletb.2016.01.020;%%

\bibitem{old1}
  L.D.~McLerran and T.~Toimela,
  {\it Photon and Dilepton Emission from the Quark--Gluon Plasma: 
  Some General Considerations,}
  Phys.\ Rev.\ D {31} (1985) 545.
  %%CITATION = doi:10.1103/PhysRevD.31.545;%%

\bibitem{old2}
  H.A.~Weldon,
  {\it Reformulation of finite-temperature dilepton production,}
  Phys.\ Rev.\ D {42} (1990) 2384.
  %%CITATION = doi:10.1103/PhysRevD.42.2384;%%

\bibitem{old3}
  C.~Gale and J.I.~Kapusta,
  {\it Vector dominance model at finite temperature,}
  Nucl.\ Phys.\ B {357} (1991) 65.
  %%CITATION = doi:10.1016/0550-3213(91)90459-B;%%

\bibitem{bps}
  R.~Baier, B.~Pire and D.~Schiff,
  {\it Dilepton production at finite temperature: 
  Perturbative treatment at order $\alpha_s$,}
  Phys.\ Rev.\ D {38} (1988) 2814.
  %%CITATION = doi:10.1103/PhysRevD.38.2814;%%

\bibitem{ggp}
  Y.~Gabellini, T.~Grandou and D.~Poizat,
  {\it Electron-positron annihilation in thermal {QCD},}
  Annals Phys.\  {202} (1990) 436.
  %%CITATION = doi:10.1016/0003-4916(90)90231-C;%%

\bibitem{aa}
  T.~Altherr and P.~Aurenche,
  {\it Finite Temperature QCD Corrections to Lepton Pair Formation
  in a Quark-Gluon Plasma,}
  Z.\ Phys.\ C {45} (1989) 99.
  %%CITATION = doi:10.1007/BF01556676;%%

\bibitem{htl}
  E.~Braaten, R.D.~Pisarski and T.C.~Yuan,
  {\it Production of Soft Dileptons in the Quark-Gluon Plasma,}
  Phys.\ Rev.\ Lett.\  {64} (1990) 2242.
  %%CITATION = doi:10.1103/PhysRevLett.64.2242;%%

\bibitem{gelis1}
  P.~Aurenche, F.~Gelis, R.~Kobes and E.~Petitgirard,
  {\it Breakdown of the hard thermal loop expansion near the light cone,}
  Z.\ Phys.\ C {75} (1997) 315
  [hep-ph/9609256].
  %%CITATION = doi:10.1007/s002880050475;%%

\bibitem{kls}
  J.I.~Kapusta, P.~Lichard and D.~Seibert,
  {\it High-energy photons from quark-gluon plasma versus hot hadronic gas,}
  Phys.\ Rev.\ D {44} (1991) 2774, 
  {\it ibid.}  {47} (1993) 4171 (E).
  %%CITATION = doi:10.1103/PhysRevD.47.4171, 10.1103/PhysRevD.44.2774;%%

\bibitem{rb}
  R.~Baier, H.~Nakkagawa, A.~Ni\'egawa and K.~Redlich,
  {\it Production rate of hard thermal photons and screening
  of quark mass singularity,}
  Z.\ Phys.\ C {53} (1992) 433.
  %%CITATION = doi:10.1007/BF01625902;%%

\bibitem{ar}
  T.~Altherr and P.V.~Ruuskanen,
  {\it Low-mass dileptons at high momenta in ultra-relativistic
  heavy-ion collisions,}
  Nucl.\ Phys.\ B {380} (1992) 377.
  %%CITATION = doi:10.1016/0550-3213(92)90249-B;%%

\bibitem{gelis2}
  P.~Aurenche, F.~Gelis, R.~Kobes and H.~Zaraket,
  {\it Bremsstrahlung and photon production in thermal QCD,}
  Phys.\ Rev.\ D {58} (1998) 085003
  [hep-ph/9804224].
  %%CITATION = doi:10.1103/PhysRevD.58.085003;%%

\bibitem{gelis3}
  P.~Aurenche, F.~Gelis and H.~Zaraket,
  {\it Landau-Pomeranchuk-Migdal effect in thermal field theory,}
  Phys.\ Rev.\ D {62} (2000) 096012
  [hep-ph/0003326].
  %%CITATION = doi:10.1103/PhysRevD.62.096012;%%

\bibitem{amy1}
  P.B.~Arnold, G.D.~Moore and L.G.~Yaffe,
  {\it Photon emission from ultrarelativistic plasmas,}
  JHEP {11} (2001) 057
  [hep-ph/0109064].
  %%CITATION = doi:10.1088/1126-6708/2001/11/057;%%

\bibitem{amy2}
  P.B.~Arnold, G.D.~Moore and L.G.~Yaffe,
  {\it Photon emission from quark gluon plasma:
  Complete leading order results,}
  JHEP {12} (2001) 009
  [hep-ph/0111107].
  %%CITATION = doi:10.1088/1126-6708/2001/12/009;%%

\bibitem{agmz}
  P.~Aurenche, F.~Gelis, G.D.~Moore and H.~Zaraket,
  {\it Landau-Pomeranchuk-Migdal resummation for dilepton production,}
  JHEP {12} (2002) 006
  [hep-ph/0211036].
  %%CITATION = doi:10.1088/1126-6708/2002/12/006;%%

\bibitem{mg}
  M.E.~Carrington, A.~Gynther and P.~Aurenche,
  {\it Energetic di-leptons from the Quark Gluon Plasma,}
  Phys.\ Rev.\ D {77} (2008) 045035
  [0711.3943].
  %%CITATION = doi:10.1103/PhysRevD.77.045035;%%

\bibitem{nlo1}
  J.~Ghiglieri {\it et al},
  %% J.~Hong, A.~Kurkela, E.~Lu, G.D.~Moore and D.~Teaney,
  {\it Next-to-leading order thermal photon production in a weakly
  coupled quark-gluon plasma,}
  JHEP {05} (2013) 010
  [1302.5970].
  %%CITATION = doi:10.1007/JHEP05(2013)010;%%

\bibitem{nlo2}
  J.~Ghiglieri and G.D.~Moore,
  {\it Low Mass Thermal Dilepton Production at NLO 
  in a Weakly Coupled Quark-Gluon Plasma,}
  JHEP {12} (2014) 029
  [1410.4203].
  %%CITATION = doi:10.1007/JHEP12(2014)029;%%

\bibitem{master} 
  M.~Laine,
  {\it Thermal 2-loop master spectral function at finite momentum,}
  JHEP {05} (2013) 083
  [1304.0202].
  %%CITATION = doi:10.1007/JHEP05(2013)083;%%

\bibitem{nlo}
  M.~Laine,
  {\it NLO thermal dilepton rate at non-zero momentum,}
  JHEP {11} (2013) 120
  [1310.0164].
  %%CITATION = doi:10.1007/JHEP11(2013)120;%%

\bibitem{interpolation}
  I.~Ghisoiu and M.~Laine,
  {\it Interpolation of hard and soft dilepton rates,}
  JHEP {10} (2014) 83
  [1407.7955].
  %%CITATION = doi:10.1007/JHEP10(2014)083;%%

\bibitem{simon}
  S.~Caron-Huot,
  {\it Asymptotics of thermal spectral functions,}
  Phys.\ Rev.\  D {79} (2009) 125009
  [0903.3958].
  %%CITATION = doi:10.1103/PhysRevD.79.125009;%%

\bibitem{cond}
  Y.~Burnier and M.~Laine,
  {\it Towards flavour diffusion coefficient and electrical conductivity
  without ultraviolet contamination,}
  Eur.\ Phys.\ J.\ C {72} (2012) 1902
  [1201.1994].
  %%CITATION = doi:10.1140/epjc/s10052-012-1902-8;%%

\bibitem{baikov}
  P.A.~Baikov, K.G.~Chetyrkin and J.H.~K\"uhn,
  {\it Order $\alpha_s^4$ QCD Corrections to $Z$ and $\tau$ Decays,}
  Phys.\ Rev.\ Lett.\  {101} (2008) 012002
  [0801.1821].
  %%CITATION = doi:10.1103/PhysRevLett.101.012002;%%

\bibitem{baikov2}
  P.A.~Baikov, K.G.~Chetyrkin, J.H.~K\"uhn and J.~Rittinger,
  {\it Adler Function, Sum Rules and Crewther Relation 
  of Order O($\alpha_s^4$): the Singlet Case,}
  Phys.\ Lett.\ B {714} (2012) 62
  [1206.1288].
  %%CITATION = doi:10.1016/j.physletb.2012.06.052;%%

\bibitem{mass}
  Y.~Burnier, M.~Laine and M.~Veps\"al\"ainen,
  {\it Heavy quark medium polarization at next-to-leading order,}
  JHEP {02} (2009) 008
  [0812.2105].
  %%CITATION = doi:10.1088/1126-6708/2009/02/008;%%

\bibitem{yannis}
  Y.~Burnier,
  {\it Quarkonium spectral function in medium at next-to-leading order
  for any quark mass,}
  Eur.\ Phys.\ J.\ C {75} (2015) 529
  [1410.1304].
  %%CITATION = doi:10.1140/epjc/s10052-015-3752-7;%%

\bibitem{rev}
  H.B.~Meyer, 
  {\it Transport Properties of the Quark-Gluon Plasma: A Lattice QCD
  Perspective,} 
  Eur.\ Phys.\ J.\ A {47} (2011) 86
  [1104.3708].
  %%CITATION = doi:10.1140/epja/i2011-11086-3;%%
 
\bibitem{photon}
  J.~Ghiglieri, O.~Kaczmarek, M.~Laine and F.~Meyer,
  {\it Lattice constraints on the thermal photon rate,}
  Phys.\ Rev.\ D {94} (2016) 016005
  [1604.07544].
  %%CITATION = doi:10.1103/PhysRevD.94.016005;%%

\bibitem{qhat}
  S.~Caron-Huot,
  {\it $O(g)$ plasma effects in jet quenching,}
  Phys.\ Rev.\ D {79} (2009) 065039
  [0811.1603].
  %%CITATION = doi:10.1103/PhysRevD.79.065039;%%

\bibitem{harvey}
  B.B.~Brandt, A.~Francis, T.~Harris, H.B.~Meyer and A.~Steinberg,
  {\it An estimate for the thermal photon rate from lattice QCD,}
  EPJ Web Conf.\  {175} (2018) 07044
  [1710.07050].
  %%CITATION = doi:10.1051/epjconf/201817507044;%%

\bibitem{screening}
  B.B.~Brandt, A.~Francis, M.~Laine and H.B.~Meyer,
  {\it A relation between screening masses and real-time rates,}
  JHEP {05} (2014) 117
  [1404.2404].
  %%CITATION = doi:10.1007/JHEP05(2014)117;%%

\bibitem{hbm}
  H.B.~Meyer,
  {\it Euclidean correlators at imaginary spatial momentum and
  their relation to the thermal photon emission rate,}
  Eur.\ Phys.\ J.\ A {54} (2018) 192
  [1807.00781].
  %%CITATION = doi:10.1140/epja/i2018-12633-0;%%

\bibitem{gj}
  G.~Jackson,
  {\it Two-loop thermal spectral functions with general kinematics,}
  1910.07552.
  %%CITATION = ARXIV:1910.07552;%%
  
\bibitem{ga}
  G.~Aarts and J.M.~Mart{\'i}nez Resco,
  {\it Continuum and lattice meson spectral functions at 
  nonzero momentum and high temperature,}
  Nucl.\ Phys.\ B {726} (2005) 93
  [hep-lat/0507004].
  %%CITATION = doi:10.1016/j.nuclphysb.2005.08.012;%%

\bibitem{dr1}
  P. Ginsparg, 
  {\it First and second order phase transitions 
   in gauge theories at finite temperature,}
  Nucl.\ Phys.\ B 170 (1980) 388.
  %%CITATION = doi:10.1016/0550-3213(80)90418-6;%%
  
\bibitem{dr2}
  T. Appelquist and R.D. Pisarski,
  {\it High-temperature Yang-Mills theories and three-dimensional 
   Quantum Chromodynamics,}
  Phys.\ Rev.\ D 23 (1981) 2305.
  %%CITATION = doi:10.1103/PhysRevD.23.2305;%%

\bibitem{generic}
  K.~Kajantie, M.~Laine, K.~Rummukainen and M.~Shaposhnikov,
  {\it Generic rules for high temperature dimensional reduction and 
  their application to the Standard Model,}
  Nucl.\ Phys.\ B {458} (1996) 90
  [hep-ph/9508379].
  %%CITATION = doi:10.1016/0550-3213(95)00549-8;%%

\bibitem{nlo_htl}
  S.~Carignano, M.E.~Carrington and J.~Soto,
  {\it The HTL Lagrangian at NLO: the photon case,}
  1909.10545.
  %%CITATION = ARXIV:1909.10545;%%

\bibitem{relativistic}
  M.~Laine,
  {\it Thermal right-handed neutrino production rate
  in the relativistic regime,}
  JHEP {08} (2013) 138
  [1307.4909].
  %%CITATION = doi:10.1007/JHEP08(2013)138;%%

\bibitem{bulk_ope}
  M.~Laine, M.~Veps\"al\"ainen and A.~Vuorinen,
  {\it Ultraviolet asymptotics of scalar and pseudoscalar
  correlators in hot Yang-Mills theory,}
  JHEP {10} (2010) 010
  [1008.3263].
  %%CITATION = doi:10.1007/JHEP10(2010)010;%%

\bibitem{sumrule}
  P.~Gubler and D.~Satow,
  {\it Finite temperature sum rules in the vector channel
  at finite momentum,}
  Phys.\ Rev.\ D {96} (2017) 114028
  [1710.02256].
  %%CITATION = doi:10.1103/PhysRevD.96.114028;%%

\bibitem{muT1}
  S.~Huang and M.~Lissia,
  {\it The Relevant scale parameter in the high temperature phase of QCD,}
  Nucl.\ Phys.\ B {438} (1995) 54
  [hep-ph/9411293].
  %%CITATION = doi:10.1016/0550-3213(95)00007-F;%%

\bibitem{muT2}
  M.~Laine and Y.~Schr\"oder,
  {\it Two-loop QCD gauge coupling at high temperatures,}
  JHEP {03} (2005) 067
  [hep-ph/0503061].
  %%CITATION = doi:10.1088/1126-6708/2005/03/067;%%

\bibitem{alph1}
  P.A.~Baikov, K.G.~Chetyrkin and J.H.~K\"uhn,
  {\it Five-Loop Running of the QCD coupling constant,}
  Phys.\ Rev.\ Lett.\  {118} (2017) 082002
  [1606.08659].
  %%CITATION = doi:10.1103/PhysRevLett.118.082002;%%

\bibitem{alph2}
  F.~Herzog {\it et al}, %% B.~Ruijl, T.~Ueda, J.A.M.~Vermaseren and A.~Vogt,
  {\it The five-loop beta function of Yang-Mills theory with fermions,}
  JHEP {02} (2017) 090
  [1701.01404].
  %%CITATION = doi:10.1007/JHEP02(2017)090;%% 

\bibitem{alph3}
  T.~Luthe, A.~Maier, P.~Marquard and Y.~Schr\"oder,
  {\it The five-loop Beta function for a general gauge group and
  anomalous dimensions beyond Feynman gauge,}
  JHEP {10} (2017) 166
  [1709.07718].
  %%CITATION = doi:10.1007/JHEP10(2017)166;%%

\bibitem{Tc}
  A.~Francis {\it et al}, %% O.~Kaczmarek, M.~Laine, T.~Neuhaus and H.~Ohno,
  {\it Critical point and scale setting in SU(3) plasma: An update,}
  Phys.\ Rev.\ D {91} (2015) 096002
  [1503.05652].
  %%CITATION = doi:10.1103/PhysRevD.91.096002;%%

\bibitem{new}
  B.B.~Brandt, M.~C\`e, A.~Francis,
  T.~Harris, H.B.~Meyer, A.~Steinberg and A.~Toniato, 
  1912.00292 and {\em work in preparation}. 
  %%CITATION = ARXIV:1912.00292;%%

\bibitem{nf2_Tc}
  B.B.~Brandt {\it et al}, 
  %% A.~Francis, H.B.~Meyer, O.~Philipsen, D.~Robaina and H.~Wittig,
  {\it On the strength of the $U_A(1)$ anomaly at the chiral phase 
  transition in $N_f=2$ QCD,}
  JHEP {12} (2016) 158
  [1608.06882].
  %%CITATION = doi:10.1007/JHEP12(2016)158;%%

\bibitem{nf2_r0}
  P.~Fritzsch {\it et al},
  %% F.~Knechtli, B.~Leder, M.~Marinkovic, 
  %% S.~Schaefer, R.~Sommer and F.~Virotta,
  {\it The strange quark mass and Lambda parameter of two flavor QCD,}
  Nucl.\ Phys.\ B {865} (2012) 397
  [1205.5380].
  %%CITATION = doi:10.1016/j.nuclphysb.2012.07.026;%%

\bibitem{nf2_Lambda}
  S.~Aoki {\it et al.} [Flavour Lattice Averaging Group],
  {\it FLAG Review 2019,}
  1902.08191.
  %%CITATION = ARXIV:1902.08191;%%

\bibitem{susc}
  A.~Vuorinen,
  {\it Quark number susceptibilities of hot QCD up to $g^6 \ln g$,}
  Phys.\ Rev.\ D {67} (2003) 074032
  [hep-ph/0212283].
  %%CITATION = doi:10.1103/PhysRevD.67.074032;%%
  
\bibitem{hydro}
  Y.~Burnier and C.~Gastaldi,
  {\it Contribution of next-to-leading order and
  Landau-Pomeranchuk-Migdal corrections to thermal dilepton emission
  in heavy-ion collisions,}
  Phys.\ Rev.\ C {93} (2016) 044902
  [1508.06978].
  %%CITATION = doi:10.1103/PhysRevC.93.044902;%%

\bibitem{landau}
  E.M.~Lifshitz and L.P.~Pitaevskii, 
  {\it Statistical Physics, Part 2}, \S88-89.
  %%CITATION = NONE;%%

\bibitem{ht}
  J.~Hong and D.~Teaney,
  {\it Spectral densities for hot QCD plasmas in a leading log approximation,}
  Phys.\ Rev.\ C {82} (2010) 044908
  [1003.0699].
  %%CITATION = doi:10.1103/PhysRevC.82.044908;%%

\bibitem{Dlatt}
  H.T.~Ding, O.~Kaczmarek and F.~Meyer,
  {\it Thermal dilepton rates and electrical conductivity
  of the QGP from the lattice,}
  Phys.\ Rev.\ D {94} (2016) 034504
  [1604.06712].
  %%CITATION = doi:10.1103/PhysRevD.94.034504;%%

\bibitem{Dpert}
  P.B.~Arnold, G.D.~Moore and L.G.~Yaffe,
  {\it Transport coefficients in high temperature gauge theories. 
  (II) Beyond leading log,}
  JHEP {05} (2003) 051
  [hep-ph/0302165].
  %%CITATION = doi:10.1088/1126-6708/2003/05/051;%%

\bibitem{shear}
  J.~Ghiglieri, G.D.~Moore and D.~Teaney,
  {\it QCD Shear Viscosity at (almost) NLO,}
  JHEP {03} (2018) 179
  [1802.09535].
  %%CITATION = doi:10.1007/JHEP03(2018)179;%%

\end{thebibliography}
\end{document}